\documentclass[11pt]{article}
\usepackage{geometry}    
\usepackage[title,titletoc,toc]{appendix}
\usepackage{hyperref}             % See geometry.pdf to learn the layout options. There are lots.
\usepackage{graphicx}
\usepackage{amssymb}
\usepackage{amsmath}
\usepackage{amsfonts}
\usepackage{epstopdf}
\usepackage[square,numbers,sort&compress]{natbib}
\newcommand{\p}{\partial}
\newcommand{\hf}{{1\over 2}}
\newcommand{\be}{\begin{equation}}
\newcommand{\br}{\begin{eqnarray}}
\newcommand{\er}{\end{eqnarray}}
\newcommand{\ee}{\end{equation}}

\newcommand{\nn}{\nonumber}

\title{A Holographic form for Wilson's RG}
\author{B. Sathiapalan\footnote{bala@imsc.res.in}\\Institute of Mathematical Sciences\\CIT Campus, Tharamani\\
Chennai 600113, India\\and\\Homi Bhabha National Institute\\Training School Complex, Anushakti Nagar\\Mumbai 400085, India\\ and\\ Hidenori Sonoda\footnote{hsonoda@kobe-u.ac.jp}\\Physics Department, Kobe University\\Kobe 657-8501, Japan}
\date{\today}                                           % Activate to display a given date or no date

\begin{document}

\maketitle
\begin{abstract}
  An attempt is made to make precise the connection between Wilson's
  RG and ``Holographic RG'' by writing Wilson's RG in a holographic
  form.  A functional formulation is given for the exact RG evolution
  of a scalar field in $d$ (flat) dimensions. It is shown that a
  change of variables maps the action to that for a scalar field in
  $AdS_{d+1}$. This provides a holographic form for Wilson's RG that
  can be called ``Holographic RG''. This mapping can only be done for
  a specific form of the cutoff function in the Exact Renormalization
  Group formalism. The notion of scale and conformal invariance in the
  presence of a {\em finite} UV cutoff is emphasized. The discussion
  is primarily about the two-point function and the Gaussian fixed
  point. Some remarks are made about nontrivial fixed points.

\end{abstract}

\newpage
\tableofcontents
\newpage

\section{Introduction}

The idea that the scale of a theory could have a geometrical
interpretation as an extra dimension is an interesting one.  One
realization of this idea occurs in the holographic or AdS/CFT
correspondence \cite{Maldacena,Polyakov,Witten1,Witten2}, where the
AdS radius coordinate naturally occurs as a scale.\footnote{Another
  instance of this was in the loop variable approach \cite{BSLV} where
  an extra string coordinate was introduced whose zero mode was dual
  to the scale. The scale also appears as a dimension in
  \cite{Witten3} in brane engineered QCD.}. This is very clear in the
AdS metric written for instance in Poincare coordinates:
\be
ds^2 = {dz^2 +\eta_{ij}dx^idx^j\over z^2} 
\ee
Here $z=0$ is the boundary of AdS space-time. This also gives rise to
the idea of the UV-IR duality, namely that the large radius region of
AdS space corresponds to the small distance region of the boundary
field theory and {\it vice versa.}
 
In a quantum (or statistical) theory the correct way to change scale
is by performing a renormalization group transformation as described
in some classic early papers \cite{Wilson,Wegner,Wilson2}. A very
convenient form for these equations was obtained in \cite{Polchinski}
and will be used in this paper\footnote{Further developments are
  reviewed in, for instance,
  \cite{MorrisERG,Becchi,Bagnuls1,Bagnuls2,Igarashi,Rosten:2010vm}}. Accordingly if we
place the boundary at $z=\epsilon$ and change $\epsilon$ this should
be equivalent to an RG transformation of the boundary theory. In the
bulk this corresponds to radial evolution using the bulk ``radial''
Hamiltonian.  This evolution has been termed the ``holographic RG''
for the boundary theory and should logically be related to the actual
RG of the boundary theory
\cite{Akhmedov,Alvarez,Girardello,Distler,Kraus,Warner,Verlinde,Boer,Faulkner,Lizana:2015hqb,Klebanov:1999tb,Heemskerk,Morris}.

There has been a lot of progress in understanding the holographic RG
flow. However the precise connection between this flow and the
Wilsonian RG flow of the boundary theory has been only to the extent
of showing that both are flows with similar properties. A precise
mathematical map is missing. This is primarily due to the fact that
the regularization of the bulk theory has not been related to the
regularization of the boundary theory \cite{Heemskerk}. Thus, while
statements are made comparing the flow of {\em renormalized}
quantities, the map between the regulated theories has not been
made. This paper attempts to fill that gap.

\subsection*{Finite Cutoff}

An important ingredient in making this connection precise is an
understanding of scale invariance and conformal invariance in the
presence of a {\em finite} UV cutoff. Usual discussions of scale and
conformal invariance is in the context of renormalized continuum field
theories where the UV cutoff is taken to infinity.  Thus for instance
N=4 Super Yang-Mills action can be shown, first of all, to be scale
and conformally invariant classically. It has no dimensionful
parameter. Quantum mechanically one has to calculate the $\beta$
functions for the dimensionless couplings. If they vanish, it is
described as a fixed point of the RG flow and is said to be scale
invariant as a quantum theory. In writing down an exact RG (ERG) for a
CFT one has to understand conformal invariance in the presence of a
cutoff. The conformal invariance of the Wilson action, and Generating
Functional in the presence of a finite cutoff has been discussed in
detail in \cite{Rosten0,Rosten1,Rosten2,Rosten3,Sonoda}. We give an
intuitive discussion here.
  
To understand scale invariance in a theory with finite cutoff, it is
best to think in terms of a theory on a lattice and look for fixed
points of the RG flow. At the fixed point the theory is scale
invariant, even while the lattice spacing is {\em finite}. As
described by Wilson \cite{Wilson}, the RG operation consists of three
steps:
\begin{enumerate}
  \item Integrate out high momentum modes (say from $\Lambda\over 2$
    to $\Lambda$). The dimensionless momenta $p\over \Lambda$ are
    integrated out from $\hf$ to $1$.
  \item Rescale the momentum so that the dimensionless momenta are
    again in units of the cutoff, which is now $\Lambda\over 2$.
   \item Do a field redefinition so that the kinetic term has the standard normalization.
\end{enumerate}

If, after these steps, the Wilson action looks the same - which means
that all dimensionless parameters in the action are the same, then we
have a fixed point. This is described as a scale invariant theory. In
fact the equation for scale invariance of the Wilson action is nothing
but the ERG equation for the fixed point \cite{Rosten1,Sonoda}.  Note
that the lattice spacing is finite. Effectively what this means is
that there is no other dimensionful parameter (such as mass) in the
theory, so that the lattice spacing drops out of all dimensionless
physical quantities.  Operationally, scale invariance then means that
if we scale all dimensionful quantities (e.g. $p_i\to \lambda p_i$)
{\em and} $\Lambda \to \lambda \Lambda$, all dimensionless quantities
that one calculates are left invariant. The subtlety here is that due
to quantum corrections, how a dimensionful quantity has to be scaled
to see the invariance, is something that is not known a priori.  The
Wilson action itself is analytic in momenta and there is no obvious
way to read of the anomalous dimension from the form of the
action. These anomalous dimensions are eigenvalues of the fixed point
equation.  On the other hand, after one has done {\em all} the
functional integration we get the effective classical 1PI action,
which should be manifestly invariant under these scale transformations
provided we include the correct anomalous dimensions.

There is one potential source of confusion. The Wilson action
$S_\Lambda[\phi]$ is obtained by starting from a bare action
$S_B[\phi]$ and integrating out modes with momenta greater than
$\Lambda$. When $\Lambda\to 0$ we have integrated out everything. In
such a situation it is more convenient to work with the generating
functional $W_\Lambda[J]$ or the 1PI effective action
$\Gamma_\Lambda [\Phi]$ \cite{Wetterich}.  For these objects $\Lambda$
is like an IR cutoff.  When $\Lambda\to0$ the
$\Gamma_{\Lambda=0}[\Phi]$ is the usual classical effective action
$\Gamma[\Phi]$. Underlying all these is a bare action defined with a
UV cutoff $\Lambda_0$ which we typically take to $\infty$ when we take
the continuum limit. Thus a priori there are always two scales in the
problem: $\Lambda_0$ and $\Lambda$. Thus even the classical effective
action $\Gamma [\Phi]$ should be thought of as having a UV cutoff
$\Lambda_0$. The bare action $S_B[\phi,\Lambda_0]$ can be thought of
as a Wilson action obtained by integrating out modes from another bare
action $S_B'[\phi,\Lambda'_0]$ with a bigger cutoff $\Lambda'_0$. In
this way the same Wilson ERG can be written for $S_B[\phi,\Lambda_0]$.
  
As discussed by Polchinski, theories that are scale invariant are
usually conformally invariant also \cite{Polchinski}.  For conformal
invariance one can write down an equation analogous to that for scale
invariance \cite{Rosten1,Sonoda}. It is expected that there exists a
Wilson action that satisfies both these conditions at the fixed point
\cite{Polchinski2,Rosten1,Sonoda}. We can easily check this for the
Gaussian fixed point.
    
We assume that there is a continuum bare action that is classically
conformally invariant (let us take the example of a free scalar field
theory). We then add an infinitesimal interaction and ask how the
generators of conformal transformation are modified along the ERG
flow. We can think of the flow as a diffeomorphism in manifold of all
possible actions. Then this flow induces in a natural way a flow for
any operator that acts on the action (via its action on the
fields). In the UV limit it reduces to the generator for the free
theory. In the IR limit it flows to some new fixed point action. The
algebra remains the same all along the flow.  Thus schematically, let
$U(t)$ describe the evolution operator for the ERG.  \be
e^{-S_\Lambda[\phi]}= U(t) e^{-S_B[\phi]} \ee Then if $T(0)$ is the
generator of scale transformations on the bare action then we can
define $T(t)$ by 
\be
T(t)e^{-S_\Lambda[\phi]}=U(t)T(0)e^{-S_B[\phi]}=U(t)T(0)U^{-1}(t)e^{-S_\Lambda[\phi]}
\ee 
Thus $T(0)$ is a symmetry of the bare action in the UV limit.  The
transformation $T(t)$ in general will be much more complicated.  It
depends on the form of the action \cite{Rosten1,Sonoda} At the fixed
point one demands that the Wilson action be invariant under
$T(t)$. This Wilson action corresponds to a nontrivial fixed point.
Furthermore if there is an algebra obeyed by $T_i(0)$, there will be
an identical algebra obeyed by $T_i(t)$.
    
The starting point in this paper is the exact RG (ERG) equations of
Wilson. The particular form used by Polchinski is very convenient and
will be used in the paper. Since the bulk field corresponds to a
current $J$ of the boundary theory, the ERG equation will be for
$W[J]$ and $J$ is assumed to couple to some field/composite operator
with well defined dimension. The leading term in the Wilson action for
this operator is a quadratic kinetic term.  To leading order in
$1\over N$ we treat the operator coupling to $J$ as a fundamental
field. Thus $W[J]$ is just $\hf J\Delta _h J$ where $\Delta _h$ is the
propagator for {\em high} momentum modes of this field --- and is cutoff
dependent. Our goal is to obtain this from a holographic formalism.

We do this in three steps:

The first step is to write the finite ERG transformation on the Wilson
action $S_\Lambda[\phi]$ as an evolution operator acting on
$e^{-S_\Lambda [\phi]}$, which we interpret as a ``wave function''.
This is a linear equation. The linearity has been noticed since Wilson
first introduced the ERG equation in \cite{Wilson}. It has been used
recently, for example, in \cite{Morris} and in \cite{Oak}.

The next step is to write this evolution operator as a functional
integral with the time direction being the scale. This makes the RG
evolution of a $d$-dimensional theory into a $(d+1)$-dimensional
theory but with a non standard action.

The third step is to make a field redefinition that restores the
action to that of a standard $AdS_{d+1}$ kinetic term. The field
redefinition turns out to be a coarse graining transformation or
equivalently an RG transformation.  This (i.e. the requirement that
the field redefinition makes it an AdS kinetic term) places
restrictions on the initial Wilson action --- on $\Delta_h$ in our case.
 
Finally one has to modify this discussion to account for the fact that
we are actually interested in the generating functional $W_\Lambda[J]$
rather than $S_\Lambda[\phi]$.

This paper is organized as follows.  In Section 2 we discuss the
mathematics of rewriting of the ERG around the Gaussian fixed point as
a functional integral in AdS.  In Section 3 we discuss conformal
invariance of the boundary theory in the presence of a finite UV
cutoff as induced by $SO(d,2)$ of the bulk theory.  In Section 4 we
make contact with standard AdS/CFT calculations.  In Section 5
composite operators are discussed in the large N
approximation. Section 6 discusses the extension to nontrivial fixed
points.   Section 7 contains a summary and some
concluding remarks.

A word about the notation: When dealing with a $d=0$ field theory
(quantum mechanics of a particle) we use ``$x$'' to denote the field,
particularly in Section 2. Generalizing this to higher dimensions, we
have used ``$x(p)$''. The conventional ``$\phi(p)$'' is also used in
some sections.

%\section{}
%\subsection{}
\section{Holographic Form for Exact Renormalization Group Equation}

In this section we deal with the ERG of a $d$-dimensional field theory and write it in terms of a functional integral in $d+1$ dimensions. Only trivial fixed points will be discussed in detail.
The field theory is defined around the Gaussian fixed point. 
A field redefinition (change of variables) casts it into a form of a scalar field theory in $AdS_{d+1}$. In this form it looks very similar
to the systems studied in AdS/CFT calculations, however at this point it is just a mathematical transformation. In particular we leave unspecified whether the bulk field is to be identified with a source in some boundary theory or with some field. This will be discussed in the next section, where we  make contact with standard AdS/CFT calculations. 
Further more in this paper only the free theory and calculation of two-point function is discussed. 
We start with a zero dimensional system ($d=0$). Generalization to higher dimensions is very simple.

\subsection{Zero Dimensional System}

We start with a zero dimensional example of Wilson's equation as rewritten by Polchinski.

If we assume that $\psi = e^{-S}$ where $S$ is the full action, the equation takes the form:
\be  \label{Wilson}
{\partial \psi\over \partial t} = -\hf \dot G{\p \over \p x}({\p\over \p x} + 2G^{-1}x)\psi
\ee
Here $G$ is a $t$ dependent function that becomes the cutoff Green function in the field theory case. It has  form
\[
G= {K({p\over \Lambda})\over p^2}
\] where $K(p)= e^{-{p^2\over \Lambda^2}} ~ or ~ e^{-{p\over \Lambda}}$ or  any form that makes the theory UV finite.

The action $S$ is assumed to have the form
\[
S= \hf G^{-1}x^2 + S_I
\]
The first term represents a ``kinetic'' term and $S_I$ is the interacting part.
Again in the field theory case there is a UV cutoff for the bare theory that we call $\Lambda_0$ and the moving cutoff $\Lambda = \Lambda _0 e^{-t}$. Thus ${\p\over \p t}= -\Lambda {\p\over \p \Lambda}$ in the field theory.
We can write an equation for the interacting part alone by defining
\[
\psi = e^{-\hf G^{-1}x^2}\psi'
\] so that $\psi'=e^{-S_I}$:
\be	\label{Polch}
 {\p \psi'\over \p t} = -\hf \dot G {\p^2 \psi'\over \p x^2}
\ee
This was the form of the equation as written by Polchinski\cite{Polchinski}. 
This is a diffusion equation. The evolution operator is clearly
\[
e^{-\int _0^Tdt~H} = e^{-\hf \int_0^T dt~\dot G {\p^2 \over \p x^2}}=e^{-\hf (G(T)-G(0)){\p^2 \over \p x^2}}\equiv e^{\hf F(T){\p^2 \over \p x^2}}
\]
Here $F(T)=G(0)-G(T)$.
The evolution operator can be calculated in the usual way to be:
\be	\label{RGkernel}
\psi'(x_f)=e^{-S_I(x_f)}= \int dx_i~e^{-\hf {(x_f-x_i)^2\over F(T)}}e^{-S_I(x_i)}
\ee
This can be rewritten in a more convenient way as 
\be	\label{RGkernel1}
\psi'(x_f)=e^{-S_I(x_f)}= \int dx~e^{-\hf {x^2\over F(T)}}e^{-S_I(x_f+x)}
\ee
Anticipating a generalization to higher dimension we can put a label $p$ in all the variables:
\[
e^{-S_I(x_p')}=\int dx_p~e^{-\hf {x_p^2\over F(p,T)}}e^{-S_I(x_p+x_p')}
\]
where
\[
F(p,T)= G(p,0)-G(p,T)={K(p/\Lambda_0)-K(p/\Lambda)\over p^2}
\]
In the field theory context this will be called $\Delta_h$ - the propagator for high energy modes, $x_p$ (or $x$) plays the role of $\phi_h$ - the high energy modes, and $x_p'$ (or $x_f$) plays the role of the low energy modes, and $x_i$ plays the role of the original field, $\phi_l+\phi_h=\phi$. Thus we have ($S_\Lambda^I$ is the interacting part of the Wilson action with moving cutoff $\Lambda$, and $S_B^I$ is the interacting part of the bare action with a fixed cutoff $\Lambda_0$.) 
\be
e^{-S_{\Lambda}^I[\phi_l]} =\int {\cal D}\phi_h e^{-\hf \phi_h{1\over \Delta_h} \phi_h -S_B^I[\phi_l+\phi_h]}
\ee 
as the analogous equation in field theory.

\subsection{Polchinski Equation as a Functional Integral}

The Euclidean action corresponding to the Hamiltonian evolution operator in \eqref{Polch} is\footnote{This can be done by first rotating to Minkowski space, obtaining the canonical Lagrangian and then rotating back to Euclidean space. See Appendix A.}:
\be	\label{action}
S= -\hf \int _0^Tdt~{1\over \dot G} (\dot x)^2
\ee
In the holographic context, this is the bulk action. This action should not be confused with the ``boundary action'' in the definition of $\psi (x)= e^{-\hf G^{-1}x^2 - S_I(x)}$. 

Thus the path integral is
\be	\label{HolRG}
\psi'(x')=\int dx~\int_{x(0)=x;x(T)=x'} {\cal D}x(t)~ e^{\hf \int _0^Tdt~{1\over \dot G} (\dot x)^2}\psi'(x)
\ee
In this form the solution to the Polchinski RG equation giving the low energy Wilson action is written as a functional integral in one higher dimension. This is a holographic form of the Polchinski RG for the interacting part of the Wilson action. The ``bulk action'' is not in a form that is recognizable as an action for a scalar field in AdS space. We will discuss this in the next sub section.

This action can be mapped to a simpler action by a change of variables: $t\to \tau(t)$:\[
{dG\over dt}= {dG\over d\tau} {d\tau\over dt}
\]
So
\[
S= -\hf \int _ {\tau(0)}^{\tau(T)}d\tau~ {1\over ({dG\over d\tau})}({dx\over d\tau})^2
\] 
Choose $G(t)=-\tau(t)$ to get
\be \label{actionG}
S=-\hf \int _ {G(0)}^{G(T)}dG~ ({dx\over dG})^2 =\hf \int _ {\tau(0)}^{\tau(T)}d\tau~ ({dx\over d\tau})^2
\ee 
Letting $\tau(0)=0, ~\tau(T)=T_E$ we get the action for a free particle   with mass, $m=1$.

Thus the path integral solution is (restoring the label $p$)
\[
\psi'(x_p',\tau(T))=\int dx_p\int {\cal D}x_p(\tau)e^{-\int_{\tau(0)}^{\tau(T)} d\tau ~\hf {\dot x_p}^2}\psi'(x_p,\tau(0))
\]
Doing the functional integral gives the same answer \eqref{RGkernel}:
\[
\psi'(x_p',\tau(T))=\int dx_p~e^{- ~\hf {(x'_p-x_p)^2\over \tau(T)-\tau(0)}}\psi'(x_p,\tau(0))
\]
\[
\psi'(x_p',T)=\int dx_p ~e^{ ~\hf {(x_p'-x_p)^2\over G(p,T)-G(p,0)}}\psi'(x_p,0)
\]
We can now obtain the solution to the original equation \eqref{Wilson}:
\[
\psi(x_p',T)= e^{-\hf G^{-1}(p,T)x_p^{'2}} \int dx_p\int {\cal D}x_p(t)e^{\int dt ~\hf {{\dot x_p}^2\over \dot G(p,t)}} e^{\hf G(p,0)^{-1}x_p^2}\psi(x_p,0)
\]
or equivalently
\be
\psi(x_p',T)= e^{-\hf G^{-1}(p,T)x_p^{'2}} \int dx_p e^{ ~\hf {(x_p'-x_p)^2\over G(p,T)-G(p,0)}}     e^{\hf G(p,0)^{-1}x_p^2}\psi(x_p,0)
\ee

The function $G(p,T)={K\over p^2}\to 0$ as $T\to \infty$. So it is useful to rescale fields\footnote{Similar redefinitions will be considered in the next sub-section.}
: Let 
\[
K^{-1}(p,T)x_p^{'2}=y_p^2
\]
Then 
\be 
\psi(y_p,T)= e^{-\hf y_p^2} \int dx_p e^{ ~\hf {(y_p \sqrt
    {K(p,T)}-x_p)^2\over G(p,T)-G(p,0)}} e^{\hf
  G(p,0)^{-1}x_p^2}\psi(x_p,0) 
\ee
We see that in the limit $T\to \infty$ the final wave function becomes
a Gaussian regardless of the initial wave function --- complete coarse
graining.

Now 
\begin{eqnarray*}
&&-\hf G^{-1}(p,T)x_p'^2+\hf G^{-1}(p,0)x_p^2 =-\hf\int _0^T dt~ {d\over dt}(G^{-1}(p,t)x_p(t)^2)\\
&&\quad=-\hf\int _0^T dt~[-{\dot G(p)\over G(p)^2}x_p(t)^2 + 2 G^{-1}(p,t) x_p(t)\dot x_p(t)]
\end{eqnarray*}
Putting this in the exponent gives
\[
\psi(x_p',T)=  \int dx_p\int {\cal D}x_p(t)e^{\int dt ~\hf \dot
  G(p)({{\dot x_p}\over \dot G(p)}-{x_p\over G(p)})^2} \psi(x_p,0) 
\]
This is the holographic form for the Polchinski RG for the full action with ``bulk action'' being
\be	\label{HolRg1}
S=-\int _0^T dt~\hf \dot G(p)({{\dot x_p}\over \dot G(p)}-{x_p\over G(p)})^2
\ee
As with \eqref{HolRG} the bulk action has a non standard kinetic
term. We rectify this in the next sub-section. We work with
\eqref{HolRG} because that is closer to what one needs in the
holographic context. This is because in the holographic context we are
interested in the evolution of the boundary action $S_I$ where $S_I$
represents a perturbation of a CFT. For the present discussion we take
this CFT to be the Gaussian (free field) theory.

\subsection{Rewriting as a scalar action in AdS}

Our starting point is the action that was used in \eqref{HolRG} (with $f=\sqrt{-\dot G}$):
\[
S=\hf\int_0^T dt~ {\dot x^2\over f^2}
\]
To get rid of the $\dot G$ we write $x= y f $. Then
\begin{eqnarray}
{\dot x^2\over f^2} &=& (\dot y + y {d \ln f\over dt})^2 = \dot y^2 +
                        y^2 ({d \ln f\over dt})^2 + 2 y \dot y {d \ln
                        f\over dt} \nn\\
&=& \dot y^2 + y^2 ({d \ln f\over dt})^2 +   \dot y^2 {d \ln f\over
    dt}\nn\\
&=& \dot y^2 + y^2 ({d \ln f\over dt})^2 -   y^2 {d^2 \ln f\over
    dt^2} +{d  \over dt}(y^2 {d \ln f\over dt})
\end{eqnarray}
The last term gives a boundary term to the action 
\be	\label{boundary1}
S_{boundary}^1 = \hf(y^2 {d \ln f\over dt})|^T_0
\ee 

Consider the term proportional to $y^2$. Let us write $z=e^t$ so that
${d\over dt}= z{d\over dz}$. We get:
\[
y^2[ (z{d\ln f \over dz})^2 - (z{d\over dz})^2 \ln f]= e^{\ln f} (z{d\over dz})^2e^{-\ln f}
\]
Let us choose $f$ to satisfy:
\[
(z{d\over dz})^2e^{-\ln f}= (z^2p^2+m^2)e^{-\ln f}
\]
Then our action becomes
\[
\int {dz\over z}~[z^2({dy\over dz})^2 +y^2(z^2p^2+m^2)]
\]
If we take an $AdS_1$ metric $ds^2= {dz^2\over z^2}$ so that
$g_{zz}={1\over z^2}$ and $\sqrt g={1\over z}$ then this can be
written as
\[
\int dz ~\sqrt g [g^{zz}(({dy_p\over dz})^2 +p^2 y_p^2) +m^2y_p^2]
\]
This is an action for a scalar field in $AdS_1$. The $p^2$ term is
actually present only in higher dimensions $AdS_{d+1}$, where it comes
from the momentum in the $d$-dimensional boundary. We have put a label
$p$ on $y$ in anticipation of the higher dimensional case where each
field will be labeled by the boundary momentum. Since the action is
quadratic the fields with different $p$ are decoupled, so we have a
simple sum over $p$ values. The only other change will be that there
will be some powers of $z$ to account for the scaling dimensions of
various quantities.

\subsection*{Equation for $f$:}

Let us take a look at the equation for $f$:
\begin{eqnarray}
\label{f} 
&&(z{d\over dz})^2e^{-\ln f}= (z^2p^2+m^2)e^{-\ln f}\nn\\
&&\implies
 [{d^2\over dz^2} + {1\over z}{d\over dz}-(p^2+{m^2\over z^2})]{1\over f}=0
\end{eqnarray}

\subsection*{Equation for $y_p$:}

The equation for $y_p$ that follows from the action is exactly the same!
\be \label{yp}
{d^2y_p\over dz^2} + {1\over z}{dy_p\over dz}-(p^2+{m^2\over z^2})y_p=0
\ee

The solutions are Bessel functions $K_m(pz)$ and $I_m(pz)$. If we do not put an IR cutoff then we should allow $z$ to become arbitrarily large. This would imply that
$K_m(pz)$ is the only acceptable solution because it has a large $z$ behavior $e^{-pz}$whereas $I_m(pz)$ blows up for large $z$.

\subsection{Obtaining Boundary Green Function}

What have we achieved so far?  By a change of variables $x=yf$, the
evolution operator for the ERG has been written as a functional
integral of an action for a scalar field, $y$, in $AdS_{0+1}$
space. This can be called the ``bulk action.'' The terminology is
taken from the AdS/CFT correspondence with which we wish to make
contact. However we remind the reader that we have not invoked any
duality conjecture. The function $f$ is related to $G$ by
$f^2=-\dot G$. Since $f$ satisfies a Bessel differential equation,
this constrains the form of $G$. (More on this below.)

Now one can use the semiclassical approximation to solve the problem. Because the
action is quadratic, this is exact. \footnote{If we have a nontrivial fixed point then the evolution operator will be more complicated.}
What we need to do is to evaluate the functional integral after specifying some suitable boundary conditions. In the semi classical approximation, this corresponds to substituting the classical solution (subject to appropriate boundary conditions)  into the bulk action. We will ignore the one loop determinant factor in what follows.

\subsubsection{Boundary Conditions}
We start with a bare action of the d-dimensional ``boundary theory''
\be
S_B = \hf xG^{-1}x + S_{I,B}(x)
\ee
(When we attempt to make contact with AdS/CFT calculations we will
need to relate this action to the perturbed boundary CFT that is of
central interest in that context. But we postpone that issue to the
next section.)  We think of the evolution operator as acting on the
wave function $\psi (t=0)=e^{-S_{I,B}(x(0))}$. Let us check that the
evolution operator obtained using this semi-classical approximation
does give the right answer.

In such a situation we need to specify:
\begin{itemize}
\item[(a)] the boundary values of $x(t)$. Let $x_i$ and $x_f$ be the
  values of the field at $t=0$ and $t=T$.  
\end{itemize}
and
\begin{itemize}
\item[(b)] since we are interested in evaluating
  $\psi(t)= e^{-S_{I,\Lambda}}$ we need to specify
  $\psi(0) = e^{-S_{I,B}(x_i)}$.
\end{itemize}
The full answer will then involve an integration over $x_i$:
\begin{eqnarray}
	\label{kernel}
\psi(x_f,T)&=&\int dx_i~\int_{x(0)=x_i,x(T)=x_f} {\cal D}x(t)~e^{-S_{bulk}[x(t)]}\psi(x_i,0)\nn\\
&=&\int dx_i~K(x_f,T;x_i,0)\psi(x_i,0)
\end{eqnarray}
We proceed to do this calculation first, in terms of $x$ and then in terms of $y$.

\subsubsection{ Boundary Action: In Terms of $x$}

We will first do the calculation in terms of $x$ --- so there is no
reference to an $AdS$ space here but we follow the same sequence of
steps:

Let us reproduce some of the earlier equations here for convenience:
Our ``bulk'' action was \eqref{action}
\be	
S= -\hf \int _0^Tdt~{1\over \dot G} (\dot x)^2
\ee
or in terms of $G$ \eqref{actionG}
\be 
S=-\hf \int _ {G(0)}^{G(T)}dG~ ({dx\over dG})^2 
\ee 

%The result of evolving with this action is given by the equation:
%\be	
%\psi'(x_f)=e^{-S_I(x_f)}= \int dx_i~e^{-\hf {(x_f-x_i)^2\over F(T)}}e^{-S_I(x_i)}
%\ee
%with $F(T)=G(0)-G(T)$

Since this is a Gaussian kernel semi classical results give the answer. So the EOM is
\[
{dx\over dG}=\textrm{constant} =b\implies x_f-x_i=b(G_f-G_i)\]
$G_f=G(T),~~G_i=G(0)$.
Substituting this into the action gives
\[
\hf x {dx\over dG}|_0^T=\hf(x_f-x_i)b=
\hf {(x_f-x_i)^2\over G(T)-G(0)}
\]

%If we happen to pick a solution where $x_f=0$ and $G(T)=0$ (as we will in an example below) we get
%\be \label{example}
%-\hf {x_i^2\over G(0)}
%\ee

This calculation gives the kernel $K(x_f,T;x_i,0)$:
\be
K(x_f,T;x_i,0)=e^{-\hf {(x_f-x_i)^2\over G(T)-G(0)}}
\ee
Inserting this in \eqref{kernel} gives:
\[
\psi (x_f,T)=e^{-S_{I,\Lambda}(x_f,T)}= \int dx_i~e^{\hf {(x_f-x_i)^2\over G(T)-G(0)}}e^{-S_{I,B}(x_i,0)}
\]
If we let $x_i-x_f\equiv x$ we can write 
\be
e^{-S_{I,\Lambda}(x_f,T)}=\int dx~e^{\hf {x^2\over
    G(T)-G(0)}}e^{-S_{I,B}(x_f+x,0)} 
\ee
In field theory notation let $\phi = \phi_h + \phi_l$ and the
propagator $\Delta = \Delta_h+\Delta_l$ where the subscripts stand as
usual for ``high'' and ``low''.  Then, $x_i \to \phi$,
$x_f \to \phi_l$ and $x \to \phi_h$. Similarly $G_i\to \Delta$ and
$G_f\to \Delta_l$ and $G_f-G_i\to -\Delta_h$. Then the equation
becomes 
\be 
e^{-S_{I,\Lambda}[\phi_l]}=\int {\cal D}\phi_h e^{-\hf
  \phi_f{1\over \Delta_h}\phi_h }e^{-S_{I,B}(\phi_h+\phi_l)} 
\ee
We obtain the standard definition of the Wilson action as expected.

\subsection*{Semiclassical Approximation}

Because the kernel is Gaussian, the $x$ integral can be done in the
semi classical approximation. This implies solving a classical
variational problem with correct boundary conditions and substituting
the solution into the action.

The variational problem gives us a boundary condition when we require
the coefficient of $\delta x$ at the boundary be zero (concentrating
on the boundary at $t=0$). This could be simple Dirichlet boundary
conditions of the form\footnote{This is equivalent to letting the
  initial wave function $\psi(x_i,0)=\delta(x_i-a_i)$.}
\[
x(0)=a_i
\]
or it could be more complicated, involving a boundary action \cite{Faulkner}: 
\[
{dx\over dG}|_0=-{\delta S_{I,B}\over \delta x_i}
\]
Let us consider some simple examples:
\begin{enumerate}
\item
$ \mathbf {x(0)=a_i}$

This gives 
\[
S_{I}(x_f,T)=-{\hf {(x_f-a_i)^2\over G(T)-G(0)}}
\]

\item
$\mathbf {S_{I,B}=kx_i}$

Let us choose $S_{I,B}(x_i)=kx_i$ ( a source term). Then we get a constraint on $b$
\[
b=-k={x_f-x_i\over G(T)-G(0)} \implies k x_i = k x_f+k^2(G(T)-G(0))
\]
This is a delta function constraint for the $x_i$ integration.
Substituting all this we get
\[
S_{I}(x_f,T)= -\hf {(x_f-x_i)^2\over G(T)-G(0)}+kx_i=\hf k^2 (G(T)-G(0)) +k x_f
\]

In the field theory analogy choose $S_{I,B} = J \phi = J(\phi_h+\phi_l)$ to get
\[
S_{I,\Lambda}[\phi_l]= -\hf J\Delta_h J + J\phi_l
\]

Thus we reproduce the Green function that results from the bare action.
\item
$\mathbf {S_{I,B}=kx_i + \hf ax_i^2}$

We obtain a boundary condition
\[
b+ax_i+k=0 \implies {x_f-x_i\over G_f-G_i} +ax_i =-k
\]
So
\[
x_i={k(G_f-G_i)+x_f\over 1-a(G_f-G_i)}
\]
Substituting this in the on shell action $S= -\hf b (x_f-x_i) +\hf a x_i^2 + kx_i$ we get
\begin{eqnarray}
S_{I,\Lambda} &=& \hf {k(G_f-G_i)+x_f\over 1-a(G_f-G_i)}[k+ax_f] + \hf kx_f\nn\\
&=&\hf k^2{(G_f-G_i)\over 1-a(G_f-G_i)} +\hf { ax_f^2\over 1-a(G_f-G_i)} +{kx_f \over 1-a(G_f-G_i)}
\end{eqnarray}

Expanding the denominator one sees that there is an obvious
interpretation in terms of tree graphs with propagators and mass
insertions.

\end{enumerate}

\subsubsection{ Boundary Action: In Terms of $y$}

We now repeat the semiclassical calculation using $y$.  We remind the
reader that in changing variables to $x=fy$ and writing the EOM as an
AdS wave equation we obtained a boundary term \eqref{boundary1} that
we need to keep track of: 
\be 
S_{boundary}^1 = \hf(y^2 {d \ln f\over
  dt})|^T_0 
\ee
{\em This is not usually done in AdS/CFT calculations, where $f$ is
  simply taken to be 1.} At low energies ${d \ln f\over dt}=0$ so it
does not affect the low energy results.

Let us turn now to the boundary term obtained by substituting the
solution of the EOM to the action:
\be 
\hf yz {dy\over dz}|_{boundary}=\hf y {dy\over dt}|^T_0=y^2 {d\ln y\over dt}|_0^T
\ee
To obtain the two-point function of the field $y$, one extracts the
coefficient of $y^2$ by dividing by $y^2$. This gives
${\dot y\over y}|_0$. (Assuming that $y(T)=0$ as $T$ becomes large for
convenience.) This is the usual AdS/CFT result.

On the other hand, adding the boundary term we get
\be  
\hf y^2 {d \ln yf\over dt}|_0^T = \hf y^2 {\dot x\over x}|_0^T
\ee
If we divide by $x^2$ to get the two-point function of $x$, we get
\[
{1\over f^2} {\dot x\over x} = {1\over x}{dx\over dG}
\]
for the exponent of the evolution operator. If we have Dirichlet boundary conditions,
(as in case 1 in the previous subsection) we get 
\[
{1\over x}{dx\over dG}|_{x=x_0}={b\over x_0}=G(0)
\]
for the two-point function, where we have assumed a solution $x=bG$, with $G_f=x_f=0$ as $T\to \infty$.

If we neglect $f$ and set it equal to 1 and we get
${\dot y\over y}|_{y=y_0}$. As mentioned before this is similar to the
answer usually quoted in AdS/CFT calculations \footnote{However
  nothing we have done so far invokes any AdS/CFT
  correspondence.}. The two calculations agree at low energies
$p\ll\Lambda$. The propagators obtained in the usual AdS/CFT
calculations also do not have the cutoff in them - they are continuum
propagators.

\subsection*{ Constraints on $f$}

We start with the observation that both $y$ and ${1\over f}$ satisfy
the same differential equation.  From $\dot x = b\dot G$ we conclude
that $x=bG + c$ (for constant $b,c$) we see that
$y=b {G\over f} +{c\over f}$. The Wronskian between $G\over f$ and
${1\over f}$ is thus
\be	\label{constr}
W[{G\over f},{1\over f}]={d\over dt}({G\over f}) {1\over f} - {G\over
  f}{d\over dt}({1\over f}) = {\dot G\over f^2} =1 
\ee
Thus let us take the solution
for ${1\over f}, y$ to be ($z=e^t$)
\br
{1\over f(t)} &=& \alpha K_m(pz) + \beta I_m(pz)\\
{G(t)\over f(t)}&=&\gamma K_m(pz)+\delta I_m(pz)
\er

Let us calculate the Wronskian:
\begin{eqnarray}
\label{W}
W[{1\over f},{G\over f}] &=& W[\alpha K_m +\beta I_m , \gamma
                             K_m+\delta I_m]=\alpha \delta
                             W[K_m,I_m]+\beta \gamma W[I_m,K_m]\nn\\
&=& \beta \gamma - \alpha \delta =-1
\end{eqnarray}
So $\alpha \delta - \beta \gamma=1$.
Thus we can write 
\be	\label{gamma}
\gamma = -{\beta\over \alpha^2+\beta^2} + a \alpha;~~~\delta = {\alpha
  \over \alpha^2+\beta^2} + a \beta 
\ee
where $a$ is arbitrary.
Given an $f$ the solution for $G\over f$ is fixed by this
relation. One can add multiples of $1\over f$ to this solution without
violating this constraint. Thus we can determine $G$ 
\be	\label{G}
G= {\gamma K_m(pz)+\delta I_m(pz)\over \alpha K_m(pz) + \beta
  I_m(pz)}={-\beta K_m(pz)+\alpha I_m(pz)\over \alpha K_m(pz) + \beta
  I_m(pz)} +a 
\ee
Thus up to a constant $a$, $G$ is fixed once $f$ is known.

Furthermore, the solution $y$ is then $y=b {G\over f} +{c\over f}$. 

\subsubsection{ Example: Disappearance of Cutoff Function} 

Let us try to fix some of the arbitrary constants: we consider the
case where the IR cutoff is zero: On physical grounds we expect that
$G(\infty)=0$ and also $x(\infty)=0$. This is equivalent to saying
that $\phi_l=0$ when $\Lambda =0$. Thus $c=0$. Thus $y= {b G\over f}$:
\be y = b {G\over f}= b(\gamma K_m(pz)+\delta I_m(pz)) \ee with
$\gamma,\delta$ given by \eqref{gamma}. There is an arbitrariness in
$G$ due to the parameter $a$ and this also shows up in $y$.

Furthermore,
since $I_m(pz)$ diverges as $z\to \infty$, the constraint that
$G(\infty)=0$ forces the coefficient,$\delta$ of $I_m(pz)$ to be zero:
\be \label{a} 
{\alpha \over \alpha^2+\beta^2} + a\beta =0 \implies a=
-{\alpha\over \beta}{1\over \alpha^2+\beta^2} \ee
This gives
\be \label{gf}
{G\over f}= {1\over \beta}K_m(pz)
\ee
and thus
\be   \label{y}
y= {b\over \beta}K_m(pz)
\ee
Finally plugging in the value for $a$ from \eqref{a} in \eqref{G} we
get
\be  \label{G11}
G= -{1\over \beta} {K_m(pz)\over \alpha K_m(pz) +\beta I_m(pz)}
\ee
Combining \eqref{gf}and \eqref{G1} we get
\be \label{feqn}
f={1\over \alpha K_m(pz) + \beta I_m(pz)}
\ee
For large $z$, $I_m(pz)\approx e^{pz}$ and $K_m(pz)\approx e^{-pz}$. Thus
\[
G(z)\approx e^{-2pz};~~~G^{-1}\approx e^{2pz}
\]
Thus we see the exponential damping at high momenta.

From \eqref{y} we see that the usual result quoted in AdS/CFT calculations 
\[
{\dot y \over y} = {\dot K_m(pz)\over K_m(pz)}
\]
 has no exponential damping at high momenta.

A simple example involving elementary functions can be obtained if we use $m=\hf$.
 Then 
both $K_\hf$ and $I_{\pm\hf}$ are solutions.
\begin{eqnarray*}
K_\hf (pz) &=& \sqrt{\pi\over 2} {e^{-pz}\over \sqrt{pz}};~~~I_\hf(pz) =\sqrt{\pi\over 2} \sinh~(pz)
\\
y&=&{b\over \beta}\sqrt{\pi\over 2} {e^{-pz}\over \sqrt{pz}};
     ~~~x={b\over -\alpha \beta -\beta^2 {1\over
     \pi}[e^{2pz}-1]}\approx e^{-2pz},~~p\to \infty\\ 
 {\dot y\over y}&=&{\dot K_\hf(pz)\over K_\hf (pz)}= [-\hf -pz]
\end{eqnarray*}
whereas \eqref{G1} gives:
\be \label{G2}
G^{-1}= -\alpha \beta - \beta^2 {I_\hf(pz)\over K_\hf(pz)}= -\alpha
\beta -\beta^2 {1\over \pi}[e^{2pz}-1] 
\ee
We see that for small momenta, the momentum dependence is linear in
both cases. For large momenta the AdS calculation involving $y$ gives
a two-point function that does not have a cutoff function. For general
$m$, the small momentum behavior has a non analyticity of the form
$p^{2m}$.  Note the difference in the asymptotic fall off of the bulk
field $y$ and the boundary field $x$. Thus for large $p$,
$x(p)\approx e^{-pz}y(p)$. This is a field redefinition. In position
space this corresponds to a convolution of the form
\[
x(u)=\int du'~g(u-u')y(u')
\]
with $g(u)\approx {1\over u^2+z^2}$ (like a ``Yukawa potential'' in
momentum space). This is a coarse graining by a function that falls of
with distance as a power - appropriate for a CFT.

\subsection{Higher dimension}

Our strategy in zero dimension was to rewrite the RG equation by doing
some field redefinitions so that it ends up looking a wave equation in
$AdS_1$. The same strategy can be repeated in higher dimensions. The
metric of $AdS_{d+1}$ is
\[
ds^2={dz^2 + \eta_{ij}du^idu^j\over z^2}~~~~i,j:0,1, ..., (d-1)
\]

The fixed point action in higher dimensions is 
\be	 
S_0 = \int _p~ \hf x_p G^{-1}(p) x_{-p}
\ee
$p$ is defined to be the Fourier conjugate of $u$. Thus
\[
x(u)= \int {d^dp\over (2\pi)^d}x(p)e^{ipu}
\]
Thus the action  
\[
\int d^du~\hf x(u)G^{-1}(u,u')x(u')=\int {d^dp\over (2\pi)^d}\hf x_p G^{-1}(p)x_{-p}
\]
The action for the functional representation of the Polchinski RG equation is thus
\be 
-\int dt \int {d^dp\over (2\pi)^d}\hf {\dot x_p \dot x_{-p}\over \dot G(p)}
\ee
Except for an integral over $p$ this is identical to the zero
dimensional case. So we introduce factors of $z^d$ in the numerator
and denominator as before and write $dt={dz\over z}$ to get
\be 
-\int dz z^{-d+1} \int {d^dp\over (2\pi)^d}\hf {\p x_p\over \p z} {\p
  x_{-p}\over \p z}{1\over z^{-d}\dot G(p)} 
\ee
In the above expression $x(u)$ is assumed to be dimensionless and thus
$x(p)$ has dimension $-d$. Thus the combination $z^{-d}\dot G(p)$ is
dimensionless. \footnote{The next section contains a detailed
  discussion of conformal dimensions and invariant actions.}  As
before we set $z^{-d}\dot G=-f^2$ which gives finally
\be 
\int dz z^{-d+1} \int {d^dp\over (2\pi)^d}\hf {\p x_p\over \p z} {\p x_{-p}\over \p z}{1\over f^2}
\ee
Again apart from factors of $z$ this is just the zero dimensional
case. This is because the evolution equation is quadratic and so the
different $p$-modes do not couple.
  
We proceed as before to define a new field by $x_p=y_p f$. The action becomes
\[
\int dz~\int _p z^{-d+1} ({\p y\over \p z} + y {\p \ln f\over \p z})^2
\]
We have suppressed the momentum indices. Thus $y^2 = y_py_{-p}$.
Performing the same manipulations as before we get
\[
\int dz~\int_p~\{z^{-d+1} ({\p y\over \p z})^2 -{\p \over \p
  z}[z^{-d+1}{d\ln f\over dz}]y^2 + z^{-d+1}y^2 ({ d\ln f\over dz})^2
+ \underbrace{{d\over dz}[z^{-d+1}y^2 {d \ln f\over
    dz}]}_{\mathrm{Boundary~term}}\} 
\]
The terms involving $f$ can be written as
\[
e^{\ln f}z^{d-1} (z^{-d+1}{d\over dz})^2e^{-\ln f}
\]
and we require as before that
\be \label{fd} 
z^{d-1} (z^{-d+1}{d\over dz})^2e^{-\ln f}= z^{-d+1}(p^2+{m^2\over z^2})e^{-\ln f}
\ee
and then the action becomes
\be \label{ads}
\int dz~\int_p\{ z^{-d+1}({\p y_p\over \p z}{\p y_{-p}\over \p z}+
z^{-d+1}(p^2+{m^2\over z^2})y_py_{-p}\} 
\ee
\eqref{fd} can be rewritten as:
\be \label{fd1}
{\p\over \p z}(z^{-d+1} {\p\over \p z} {1\over f}) = z^{-d+1} (p^2+{m^2\over z^2}){1\over f}
\ee
As in the zero dimensional case the equation for $y_p$ obtained from \eqref{ads} is exactly the same.

The solutions are $z^{d\over 2}K_\nu (pz)$ and $z^{d\over 2}I_\nu (pz)$ where $\nu^2=m^2+{d^2\over 4}$. 

\subsection{Constraints on $f$}

\eqref{constr} becomes
\be
W[{G\over f},{1\over f}]={d\over dt}({G\over f}) {1\over f} - {G\over f}{d\over dt}({1\over f}) = {\dot G\over f^2} =z^{d}
\ee
\eqref{W} becomes
\begin{eqnarray}
W[{1\over f},{G\over f}] &=& W[\alpha z^{d\over 2}K_\nu +\beta
z^{d\over 2} I_\nu , \gamma z^{d\over 2}K_\nu +\delta z^{d\over
  2}I_\nu]\nonumber\\
&=&\alpha \delta z^d W[K_\nu,I_\nu]+\beta \gamma z^dW[I_\nu,K_\nu]
\nonumber\\
&=& z^d(\beta \gamma - \alpha \delta) =-z^d
\end{eqnarray}
This leads to the same relation
\[
\alpha \delta - \beta \gamma =1
\]
Thus the expression for $G$ remains the same as \eqref{G}
\be	\label{G1}
G= {\gamma K_\nu(pz)+\delta I_\nu(pz)\over \alpha K_\nu(pz) + \beta I_\nu(pz)}={-\beta K_\nu(pz)+\alpha I_\nu(pz)\over \alpha K_\nu(pz) + \beta I_\nu(pz)} +a
\ee

\subsection*{Example:}

In particular one can consider the example with $\nu=\hf$ that was studied in an earlier subsection. Let us choose $d=2$ for concreteness. This would mean $m^2+{d^2\over 4}= \nu^2 \implies m^2=-{3\over 4}$. \footnote{Note that in AdS values of $m^2$ greater than $-{d^2\over 4}$ are consistent.}  

In \eqref{f} if we let $p\to 0$ we get $f= {1\over \alpha K_\hf (pz)} \approx {\sqrt{pz}\over \alpha}$. Thus if we choose $\alpha = \sqrt p$, $f$ goes to a constant at small values of momenta, which means $x$ and $y$ have the same low energy behavior. Thus let us choose $\alpha =\beta = \sqrt p$. Then
\[
G^{-1} = {\beta^2\over \pi}e^{2pz}
\]
and thus
\be \label{G22}
G\approx{e^{-2pz}\over p}
\ee

We will see in the next section, the same Green function emerge from a natural choice of regularization in the AdS bulk.

\section{Conformal Invariance of Actions}
\label{1}

The function, $f$, has been chosen so that the bulk equations have the standard $AdS_{d+1}$ form and therefore one expects an SO(d,2) symmetry for the action - which is conformal invariance in $d$ dimensions. 

The conformal transformation in these coordinates are given in the Appendix. The main new feature is that the coordinate $z$, which from the $d$ dimensional point of view is like a cutoff scale, also transforms under the transformations. From the point of view of the boundary theory we should understand this as conformal invariance in the presence of a cutoff scale. As discussed in the introduction, the conformal invariance of the Wilson action has been studied in a general way in \cite{Rosten1,Sonoda}. The main message is that these are conformally invariant but the transformation law for fields is modified due to the presence of the cutoff function in the Wilson action. The modification also depends on the Wilson action itself. In this section we look at this issue in the context of AdS/CFT which provides naturally  a very specific form for the cutoff function and also a specific form for conformal transformations induced from the SO(d,2) symmetry of the bulk theory, which in turn is induced by the  isometry of $AdS_{d+1}$.

Some of the issues that arise when checking conformal invariance, especially in the presence of a finite cutoff, are discussed below. Some details are given in the Appendix.

\subsection{An Example of Regulated Green function in AdS space}

In the AdS/CFT correspondence the Green function of the boundary theory is induced
in a natural way from that of the bulk theory. Thus if the bulk Green function is finite as $x\to x'$ (here we use the usual notation $x,x'$ for the boundary coordinates and $z,z'$ for the radial coordinate in the AdS bulk) the boundary Green function will also be so. We consider an example below with $d=2$.

$AdS_{3}$ space can be defined by the equation \eqref{ads} reproduced below.
\be 
-(y^0)^2+(y^1)^2+(y^2)^2-(y^4)^2=-1
\ee
The invariant $\xi=y.y'= {(x-x')^2 + z^2 + z'^2\over 2zz'}$ is related to  the geodesic distance $D$ by
\[
\eta\equiv\xi-1=  {(x-x')^2 + (z-z'^2)\over 2zz'}= \cosh (D) -1
\]
Thus for small $D$ we have 
\[
{(x-x')^2 + (z-z'^2)\over 2zz'}={D^2\over 2}
\]

Thus consider the function on the boundary:
\[
G(x-x')=({1\over (x-x')^2 + (z-z'^2)})^\alpha
\]
Let us choose $\alpha =\hf$.
Its Fourier transform is ($\nu = {d\over 2}-\alpha =\hf$)
\begin{eqnarray}
G(p) &=& 2 K_\nu (p(z-z'))({p\over z-z'})^{-\hf }\nn\\
&=&2 \sqrt{\pi\over 2} {e^{-p(z-z')}\over p}
\end{eqnarray}
This is precisely of the form given by \eqref{G22}.

In the free (Gaussian fixed point) theory with this $G$, the two-point
function is thus up to normalization (\eqref{G22} is obtained by
replacing $z$ by $2z$)
\be \label{lattice}
\langle \phi(p,z) \phi(-p,0)\rangle = {e^{-pz}\over p}
\ee
The low energy version (or equivalently the continuum limit) is
\be \label{cont}
\langle \phi(p,z) \phi(-p,0)\rangle = {1\over p}
\ee
When one acts on correlation functions, the conformal transformations
are in the end acting on {\em functions}. The special conformal
generators are, for an object of dimension $\Delta$, and acting on
functions of $p^2$,
\be  \label{ct}
C_\mu = C_\mu^1+C_\mu ^2 + C_\mu^3
\ee
with
\begin{subequations}
 \label{ct1}
\begin{eqnarray}
C^1_\mu &=& p_\mu[{d^2\over dp^2}+ {d+1-2\Delta\over p}{d\over dp}] \\ 
C^2_\mu&=& -2z { d\over dz}{d\over dp^\mu}\\
C^3_\mu&=& z^2p_\mu
\end{eqnarray}
\end{subequations}
The conformal transformations of correlation functions in a continuum
CFT ($z=0$) are given by $C_\mu^1$ in \eqref{ct1}. It is easy to see
that \eqref{cont} is invariant under this when the scaling dimension
of $\phi(x)$, $\Delta=\hf$. This corresponds to a scale and
conformally invariant action 
\be 
S_{\mathrm{continuum}} =\int
d^2p~\phi(p)p\phi(-p) 
\ee
The transformation $C_\mu^1+C_\mu^2+C_\mu^3$ given in \eqref{ct1}
leave the two-point function \eqref{lattice} invariant. This involves
transformations of the cutoff $z$. This is an example of conformal
invariance of a fixed point theory on a lattice.

The lattice action (by which we mean an action with a finite UV cutoff) is of the form
\be
S_{lattice}= \int d^2p~\phi(p)p K^{-1}(p)\phi(-p)=\int d^2p~\phi(p)pe^{pz}\phi(-p)
\ee
This can also be thought of as a Wilson action of a continuum
theory. With this interpretation the invariance of actions of this
type are discussed in detail by \cite{Rosten1,Sonoda}.  In summary it
involves changing the conformal transformation of $\phi$ from the
continuum form to something that involves the cutoff.
\be
\delta \phi = \epsilon K(p)C_\mu^1 \phi 
\ee
This is valid for arbitrary $K(p)$ and not just the form given here
involving Bessel functions. For an interacting theory the
transformation also depends on the Wilson action itself. The specialty
of the form involving Bessel functions obtained in Section 2 by
demanding an AdS isometry, is that the transformations have a simple
form given in \eqref{ct1}.

\subsection{Finite range in $z$ and boundary terms}
\label{2}
Let us consider the form of some invariant actions.

If the field $\phi(x)$ is dimensionless then
\be   \label{ci}
S_1=\hf \int d^dx~\int {dz\over z}~z^{-d+2} [\p_z \phi (x,z)\p_z \phi(x,z) + \p_i \phi(x,z)\p_i \phi(x,z)]
\ee
is scale and conformally invariant.
More generally the action (we write it in momentum space)
\be  \label{cip}
S_2 = \hf \int {d^dp\over (2\pi)^d}~\int {dz\over
  z}~z^{-d+2\Delta}[z\p_z \phi (p,z)z\p_z \phi(-p,z) + (p^2z^2+m^2)
\phi(p,z) \phi(-p,z)] 
\ee
is scale and conformally invariant.  Here $\Delta$ is the conformal
dimension of $\phi(x)$. (Conformal dimension of $\phi(p)$ is then
$\Delta -d$.)

Actually, when the integral over $z$ is over a finite range
$\epsilon_1$ to $\epsilon_2$ (as is usually the case) the invariance
of the above actions is modulo boundary terms, as can be checked by
the conformal transformation rules given in the Appendix.  Thus if
\be
S_1=\hf \int d^dx~\int_{\epsilon_1}^{\epsilon_2} {dz\over z}~z^{-d+2}
[\p_z \phi (x,z)\p_z \phi(x,z) + \p_i \phi(x,z)\p_i \phi(x,z)]\,, 
\ee
the boundary term under dilatations of $S_1$ is (see Appendix)
\be  \label{d-b}
\Delta_{dil} S_{boundary~term}=-\{\hf\int d^dx z^{-d+2}[\p_z \phi \p_z
\phi + \p_k \phi \p_k \phi ]\}|_{\epsilon_1}^{\epsilon_2}  
\ee
Similarly, the boundary term under conformal transformation $C_\mu^i$
is
\be  \label{sc-b}
\Delta_{conf} S_{boundary~term}=-\{\hf\int d^dx z^{-d+2}x^i[\p_z \phi
\p_z \phi + \p_k \phi \p_k \phi ]\}|_{\epsilon_1}^{\epsilon_2}  
\ee
Both these terms can be canceled by transforming the end points of the action by the respective transformations.
Thus under scale transformations we see:
\be
-\sum _{a=1,2}\epsilon_a {d\over d\epsilon_a} S_1 +\Delta_{dil} S_{boundary~term}=0
\ee
and under conformal transformations:
\be
-2x^i\sum _{a=1,2}\epsilon_a {d\over d\epsilon_a}S_1+\Delta_{conf} S_{boundary~term}=0
\ee
Intuitively, if a scale invariant system is put in a {\em finite} box
one expects the scale transformations to be violated unless the box is
also stretched. Thus one can give a modified prescription for scale
invariance that stretches the box also. This is again similar to the
rescaling that accompanies an RG transformation, when looking for a
fixed point. What is fixed is an action where {\em dimensionful
  quantities are measured in units of the lattice spacing}. In the
problem above $\epsilon$ plays the role of lattice spacing in the
boundary theory and sets the scale. The extra complication is that one
of them is a UV cutoff and the other an IR cutoff, and both have to be
rescaled together.

\subsection{Rescaling fields with cutoff scale or $z$}

In studying RG it is often convenient to rescale operators by powers
of the cutoff $\Lambda$ to make them dimensionless. This is equivalent
to multiplying by powers of $z$ (inverse cutoff or lattice spacing) in
the present context.  Note that it is the combination
$\Delta + z {d\over dz}$ that occurs in the scale transformation as
well as special conformal transformation (see Appendix). So if $\phi'$
has dimensions $\Delta$, and $\phi'=z^b\phi$, then $\phi$ has
dimension $\Delta+b$:
\be  \label{dil}
\delta \phi' = (\Delta + z {d\over dz})\phi'  =(\Delta + z {d\over
  dz})z^b \phi = z^b (\Delta +b +z {d\over dz})\phi = z^b \delta \phi=
\delta (z^b \phi) 
\ee 

Let us start with the above conformally invariant action for a dimensionless field $\phi'$, with $\Delta =0$.
In momentum space this is
\be
\int {d^dp\over (2\pi)^d}~\int {dz\over z}~z^{-d}[z\p_z \phi
(p,z)z\p_z \phi(-p,z) + (p^2z^2+m'^2) \phi(p,z) \phi(-p,z)] 
\ee
Let us define $\phi' = z^b \phi$. If we substitute we obtain the
action
\be
\int {d^dp\over (2\pi)^d}~\int {dz\over z}~z^{-d+2b}[z\p_z \phi
(p,z)z\p_z \phi(-p,z) + (p^2z^2+m^2) \phi(p,z) \phi(-p,z)] 
\ee
where $m^2= m'^2 +bd -b^2$. Note that this is in accordance with
\eqref{cip} because the dimension of $\phi$ is $\Delta =b$ in this
case.

This kind of rescaling is done for instance when writing the RG
equations for a fixed point \cite{Wilson}. Thus we see that the
canonical AdS form of the action for a scalar field is in terms of the
rescaled dimensionless field. One can further rescale $x^i$ to a
dimensionless coordinate $\bar x^i$ as is done when looking for a
fixed point action, and then all powers of $z$ disappear from the
action.

In the next section we use the formalism of this section to elucidate
the connection between Wilson's RG and holographic RG.

\section{Making Contact with AdS/CFT}

The typical problem discussed in the context of AdS/CFT correspondence
is the following: There is a $d$-dimensional CFT. The strict
definition (when a cutoff is finite) would be that this corresponds to
a field theory at the fixed point. But at low energies, it is
sufficient to think of a field theory on the critical surface. This
field theory is perturbed by the addition of a term of the form $\int
d^dx~ J(x) O(x)$. $O(x)$ is some operator in the CFT. One is
interested in the full generating functional $Z[J]\equiv
e^{W[J]}$. One may also be interested in $Z_\Lambda[J] \equiv
e^{W_\Lambda [J]}$ where $\Lambda$ is an IR cutoff. When $\Lambda\to 0$ one recovers $Z[J]$. One reasonable way to do this is to write down an RG equation for $Z_\Lambda[J]$ and solve it with some initial condition specified at large $\Lambda$. This may be obtained in perturbation theory. Thus what one needs in this approach is an ERG equation for $Z_\Lambda[J]$ or $W_\Lambda[J]$.

The AdS/CFT duality conjecture states that $Z[J]$ can be calculated in a $d+1$ dimensional theory,  ``bulk theory'' where the space-time is $AdS_{d+1}$. $J(x)$ is elevated to a field in the bulk theory, $J(x,z)$ such that $J(0,x)=J(x)$ is the boundary value. Then $Z[J(0)]$ is given by the action of the bulk theory evaluated on a solution of the equations of motion. The solution is required to satisfy the boundary condition that $J(0,x)=J(x)$.

Furthermore if the boundary is placed at $z=\epsilon$ rather than at zero, $1\over \epsilon$ can be thought of as a UV cutoff for the boundary theory. The evolution of $J(x,\epsilon)$ and $W[J(x,\epsilon)]$ with $\epsilon$ should be thought of as a renormalization group evolution - ``holographic RG'' of the generating functional. But the evolution of $J(x,z)$ is given by the equations of motion of the bulk field in $AdS_{d+1}$ space.

If this AdS/CFT conjecture is correct, the ERG for $W_\Lambda[J]$ should be writable as an equation for a field $J(x,z)$ in AdS space.
This is almost what we have shown in Section 2. The only difference is that the ERG equation considered in Section 2 was for the Wilson action $S_\Lambda[\phi_l(x)]$ not $W_\Lambda[J]$. 

Actually, for the so called ``alternate quantization'' (see for instance \cite{Faulkner}) the boundary value of bulk field is interpreted as the effective classical field. So what we need is $\Gamma_\Lambda[\Phi]$. For external momenta lower than $\Lambda$, $S_\Lambda[\phi]$ is the same as $\Gamma_\Lambda[\phi]$, since the non-1PI graphs do not satisfy the momentum conservation constraint. So the discussion in Section 2 can be directly applied for this case. But since the more general situation is the ``standard quantization,'' we turn to that below.

In the simplest situation discussed in the last section, obtaining an equation for $W[J]$ is not a problem. This involves a Gaussian CFT and restriction to two-point functions. If $\phi$ is the elementary scalar field of the boundary CFT, and the operator $O$ is taken to be 
$\phi$ itself, then the ERG equation for $W_\Lambda[J]$ has the same (Polchinski) form as that for  $S_{I,\Lambda}[\phi_l]$. Thus indeed all the calculations of Sec 2 can be applied here. Thus the ERG equation for $W_\Lambda[J]$ does indeed have an AdS form. So in this restricted situation we have demonstrated that Wilsonian RG is indeed the same as holographic RG.

One useful check is the following: The ERG equation in general is for $W_\Lambda[J]$. As long as $\Lambda$ is finite $W_\Lambda[J]$ should be analytic at zero momentum. The non-analyticity that is usually seen due to zero momentum modes should be absent. We verify this in  Section 4.2.

The next complication is that $O$ is typically a composite operator,
such as $\phi^2$ or $\phi^4$. In that case the ERG equation for
$W_\Lambda [J]$ appears to be much more complicated. However one can
make progress if one makes a large $N$ approximation. In this
approximation higher point correlators are down by powers of $N$ and
only the two-point function is finite. In this situation  one can
replace the composite operator by an elementary field using Lagrange
multiplier fields. The ERG for $W_\Lambda [J]$ again has the Polchinski form and all the calculations go through exactly as above. This will be described in the next section.

With all this, as far as the Gaussian fixed point is concerned, and within a $1\over N$ expansion, for all two-point functions of elementary or composite fields, we have demonstrated that the holographic RG is just the Wilson RG - as long as the momenta are smaller than the UV cutoff.   

What remains is to understand nontrivial fixed points. This will be discussed in Section 6.

\subsection{ERG equation for $W_\Lambda [J]$}

The generating functional is given by (using standard field theory
notation)
\be \label{zjl}
Z_\Lambda [J]=e^{W_\Lambda[J]}=\int {\cal D}\phi_h ~e^{-\hf
  \phi_h{1\over \Delta_h}\phi_h - S_{\Lambda_0}^I [\phi_h+ \Delta_h J]
  + \frac{1}{2} J \Delta_h J}
\ee
where $\Delta_h = G(p,\Lambda_0)-G(p,\Lambda)= \left(K(p,\Lambda_0)-K(p,\Lambda)\right)/p^2$ \footnote{For more general case $p^2$ will be replaced by some power of $p$.} is the high energy propagator and $\phi_h$ is the high energy field.
We are interested in $W_\Lambda [J]$. One can
derive the following ERG for $W_\Lambda [J]$ (derived in the Appendix
for completeness): 

\be	\label{ERGW}
{dW_\Lambda \over dt}= - \hf \dot {(\Delta _h^{-1})}[({\p W_\Lambda\over \p J})^2 + {\p^2W_\Lambda\over \p J^2}]
\ee

Clearly this is almost identical to the Polchinski equation for $S_{I,\Lambda}$ that we have been using. The only difference is that $\dot \Delta_h$ is replaced by $\dot \Delta _h^{-1}$. Thus the field $x$ in Section 2 stands for $J$. $y$ is then $J\over f$ for a suitable $f$. The quantity $G$ calculated in Section 2 is to be thought of as $\Delta_h^{-1}$.

One complication here is that $\Delta_h(0)\to \infty$ as $t\to 0$. So the external source $J(p,t)$ must be renormalized suitably such that $J(p,0)$ is finite and is the external source of the boundary field theory.

\subsection{On Shell Action}

The action for $J(p,t)$ is
\be 	\label{JA}
\hf \int_0^T dt~ {\dot J^2\over \dot {(\Delta _h^{-1})}}
\ee
The EOM is
\[
{\dot J\over \dot {(\Delta _h^{-1})}}= \textrm{const}
\]
The simplest solution satisfying this is
\be	\label{js}
J(t) = \textrm{const}~ \Delta_h(t)^{-1} \implies J(t)\Delta_h(t)= \textrm{const} = j \Delta_h(\epsilon)
\ee
where we have normalized so that $J(\epsilon)=j$ for some small $\epsilon$. We eventually take $\epsilon =0$, which corresponds to $\Lambda=\Lambda_0$ which means $\Delta_h=0$ - which is the reason for factoring out $\Delta_h(\epsilon)$ in \eqref{js}.

If we now plug this solution into the action we get
\be
\hf{J\dot J\over \dot {(\Delta _h^{-1})}}|^T_\epsilon = \hf(J(T)-J(0)) j\Delta_h(\epsilon)= \hf J(T) \Delta_h(T)J(T)- \hf J(\epsilon)  \Delta_h(\epsilon)J(\epsilon)
\ee
Acting on the initial state wave function, which has $\Lambda=\Lambda_0e^{-\epsilon}$: 
\[
\psi (\epsilon)= e^{W_\Lambda[J]}=e^{\hf J(\epsilon)  \Delta_h(\epsilon)J(\epsilon)}
\]
one obtains 
\be
\psi(T)=e^{\hf J(T)  \Delta_h(T)J(T)}
\ee
as expected at this order.

If one wants to go beyond the Gaussian approximation, one has to include in the starting wave function, cubic and higher order (in $J$) terms in $W[J]$. This is what is normally done in ERG calculations. However in principle one could imagine keeping the initial wave function the same and modifying the ERG evolution action instead. This is what happens in the holographic context.  It is not clear what this means physically for the renormalization group. We discuss this briefly in Sections 6 where non trivial fixed points are discussed.

Finally it should be mentioned that in a certain range of dimensions
of operators, the bulk field can be identified with the expectation
value of the operator rather than with the source. This has been
called ``alternate quantization''. In this case the calculations of
Section 2 are directly applicable. One can use the Polchinski RG for
the Wilson action itself rather than $W[J]$.

\subsection{Implementing a finite $\Lambda$}

One of the inputs into the definition of $W_\Lambda[J]$ is the IR
cutoff $\Lambda$. Thus only momentum modes between $\Lambda$ and
$\Lambda_0$ have been integrated out. Thus the generating functional
obtained should be analytic at $p=0$ where $p$ is the label in
$J(p)$. \footnote{This same question could have been discussed in the
  last section but we chose to keep it for later.} Thus in this
section we focus a bit more on the precise functional form of the
function $G$ or $\Delta_h$.  $\Delta_h$ is of the form
${K(p,\Lambda_0)-K(p,\Lambda)\over p^\delta}$. Typically from AdS/CFT
calculation we expect that $\delta$ can be any real number - to match
the dimension of the operator - which could be essentially
anything. Thus for instance in the Sine-Gordon theory the operator
$\cos (\beta \phi)$ has dimension $\beta^2\over 2$. For an ordinary
scalar field $AdS/CFT$ calculation can give propagators like
$1\over p^2$. The power of $p$ depends on the parameter $m$ in
\eqref{ads}.

However,  when $\Lambda \neq0$ the propagator, $\Delta_h$, has to be analytic at $p=0$. If we are working with the Wilson action, rather than the generating functional, the same argument holds and one concludes that the Wilson action has to be analytic at $p=0$.
Thus from $W_\Lambda[J]$ we should recover the high energy propagator, not the full propagator. 

 In the AdS calculation this requires imposing an IR cutoff . This means we need boundary conditions, at finite $z$ - not at $z=\infty$. If we pick $z=\infty$ we are forced to pick $K_m(pz)$ as our solution. Otherwise we get a linear combination of $K_m(pz)$ and $I_m(pz)$.

The procedure in Section 2 to obtain the Green function is to solve the bulk field equation and evaluate the on-shell action. Solving the field equation in the bulk involves obtaining the Green function of the bulk theory.
We call it $\cal G$. 
We thus proceed to calculate $\cal G$ and thence the boundary to bulk Green function with Dirichlet boundary conditions at $z=\epsilon$ and $z=l$. 
The differential equation is
\be
\p_z^2 {\cal G}(z,z')-{1\over z}\p_z {\cal G}(z,z')-(p^2z^2+m^2){\cal G}(z,z') = z\delta(z-z')
\ee
Thus the conditions on the Green function are
\[
{\cal G}(l,z')=0={\cal G} (\epsilon,z')
\]
and the matching conditions at $z'$ i.e.
${\cal G}(z,z')$  for $z<z'$ and $ {\cal G}(z,z')$ for $z>z'$ agree at $z=z'$. Furthermore
\[
\int_{z'-\epsilon}^{z'+\epsilon}dz~[\p_z^2  {\cal G} -{1\over z}\p_z {\cal G} -(p^2z^2+m^2) {\cal G}]=z'
\]

The solution is of the form (we have picked $d=2$ for concreteness and $\nu=1+m^2$):
\br
 {\cal G}(z,z') &=& A(z')z K_\nu(pz) + B(z')zI_\nu(pz) ~~~~~z>z'\nonumber \\
           &=&C(z') zI_\nu (pz) +D(z')z K_\nu(pz) ~~~~~z<z'
\er
If we define
\[
F[x,y]=I_\nu(x)K_\nu(y)-K_\nu(x)I_\nu(y)
\]
then the solution is written compactly:
\begin{subequations}
\begin{eqnarray}
A&=&-z'I_\nu(l){F[pz',p\epsilon]\over F[pl,p\epsilon]}\\
B&=&z'K_\nu(l){F[pz',p\epsilon]\over F[pl,p\epsilon]}\\
C&=&z'K_\nu(\epsilon){F[pz',p\epsilon]\over F[pl,p\epsilon]}\\
D&=&-z'I_\nu(\epsilon){F[pz',p\epsilon]\over F[pl,p\epsilon]}
\end{eqnarray}
\end{subequations}
The boundary (at $ z=\epsilon$) to bulk propagator is
$K(z,\epsilon)=\sqrt g g^{zz}\p_{z'} {\cal G}(z,z')|_{z'=\epsilon}$:
\be
K(z,\epsilon)=-{z\over \epsilon} {F[pl,pz]\over F[pl,p\epsilon]}
\ee
Note that when $l\to \infty$ it reduces to the well known form
$-{z\over \epsilon} {K_\nu(pz)\over K_\nu(p\epsilon)}$.

The boundary Green function is thus given by
\[
G(p)=\p_z K(z,\epsilon)|_{\epsilon} = {\p_zK_\nu(pz)|_\epsilon \over K_\nu(p\epsilon)}-{1\over \epsilon}
\]

Let us obtain the dependence on $p$ (for small $p$)  using the properties:
\begin{subequations}
\begin{eqnarray}
{dK_\nu(x)\over dx}&=& -K_{\nu+1} + {\nu\over x}K_\nu(x)\\
K_\nu(z)&=&{\pi\over 2 \sin (\nu \pi)}[I_\nu(z) - I_{-\nu}(z)]\\
I_\nu(z)&=&({z\over 2})^\nu  {1\over \Gamma(1+\nu)} + O(z^{2+\nu})\\
K_\nu(z)&=&{\pi\over 2 \sin (\nu \pi)}\left[({z\over 2})^{-\nu} {1\over
            \Gamma(1-\nu)} + O(z^{2-\nu}) \right.\nn\\
&&\qquad\qquad\quad\left. -({z\over 2})^{\nu} {1\over
            \Gamma(1+\nu)}+O(z^{\nu+2})\right] 
\end{eqnarray}
\end{subequations}
The answer is that $G(p)$ (for small $p$), has the structure:
\[
G(p)\approx 1+O(p^2) + p^{2\nu}(1+O(p^2))
\]
Since the analytic pieces can easily be canceled by counterterms, the
only significant part is the non analytic term $p^{2\nu}$. 

We can repeat the exercise for the case of $l$ being finite. 
\be
{F[pl,pz]\over F[pl,p\epsilon]}=
{I_\nu(pl)K_\nu(pz)-I_\nu(pz)K_\nu(pl)\over
  I_\nu(pl)K_\nu(p\epsilon)-I_\nu(p\epsilon)K_\nu(pl)} 
\ee
Consider the numerator, which can be written as:
\begin{eqnarray*}
&&{\pi\over 2 \sin (\nu \pi)}(I_\nu(pl)[I_\nu(pz) -
I_{-\nu}(pz)]-I_\nu(pz)[I_\nu(pl) - I_{-\nu}(pl)])\\
&=& {\pi\over 2 \sin (\nu \pi)}(I_\nu(pl)I_{-\nu}(pz)-I_{-\nu}(pl)I_\nu(pz))
\end{eqnarray*}
Using the fact that $I_\nu(pz)$ has the structure $p^\nu (1+O(p^2)$ we
can see that the terms non analytic in $p$ have canceled and we are
left with the structure:
\be
F[l,z]\approx 1+ O(p^2)
\ee
Thus we see that the Green function has no non analytic part and is also finite as $p\to 0$. 

For later use we also give the 
the boundary (at $z=l$) to bulk propagator: 
\be
K(z,l)={z\over l} {F[z,\epsilon]\over F[l,\epsilon]}
\ee
We can also see that (as explained in Section 2) in this expression
there is no damping at high $p$, i.e. $\Lambda_0=\infty$. 

\section{Composite Fields}

The typical situation in an AdS/CFT calculation is the following: We start with a CFT. Add a perturbation 
\[
\Delta S = \int d^dx~ J(x)O(x)
\]
where $O(x)$ is a composite operator with some fixed dimension. We
would like to calculate correlation functions for $O(x)$ or more
generally calculate the generating functional $Z[J]$. So far in this
paper we have been considering the case where $O(x)$ is just an
elementary scalar field used in defining the action. We took the CFT
to be the simple Gaussian theory. But even while maintaining the
simplicity of the Gaussian CFT, we can consider what happens when
$O(x)$ is some composite such as $\phi^2(x)$ or $\phi^4(x)$. This is
harder to do because the ERG equation for $J$ is no longer the simple
Polchinski form of the last section. To circumvent this problem we
invoke a further simplification: we assume that there are $N^2$ scalar
fields with $N$ large and introduce, via a Lagrange multiplier field,
an elementary field to replace the composite operator. The elementary
scalar field has nontrivial higher point correlators, but these are
accompanied by factors of $1/N$. So in the leading large $N$ limit we
are left with the two-point function which can be calculated using the
procedure of the previous section. In the next order in $1/N$ we need
to worry about the three point function but this can be done using
perturbation theory.

\subsection{Matrix model}

Let us assume that our scalar field is a matrix $\Phi ^a_b;~~a,b:1,
\cdots, N$. Assume the kinetic term
\be
N \Phi^a_{~b} (\Delta^{-1})_{ac}^{bd}\Phi_{~d}^c = N \Phi^a_{~b} \Box
/K(\Box/\Lambda^2) \Phi _{~d}^c\delta_a^d\delta_c^b 
\ee
The propagator $\Delta$ is
\[
\Delta (p)= {\delta_a^d\delta_c^b\over N}{K({p^2\over \Lambda^2})\over p^2}
\]

We see that the composite operator $Tr\phi^2$ is properly normalized:
\[
 \langle Tr\Phi^2(x) Tr\phi^2(y) \rangle\approx {1\over N^2}\times N^2 (\Delta(x-y))^2=O(1)
\]
But
\[
\langle Tr\Phi^2 (x) Tr\Phi^2 (y) Tr\Phi^2 (z) \rangle \approx {1\over
  N^3}\times N^2 (\Delta(x-y)\Delta(x-z)\Delta(z-y)) =O({1\over N}) 
\]
In general $\langle (Tr \Phi^2)^n\rangle \approx {1\over
  N^{n-2}}$.
Thus we expect that if we introduce an elementary scalar field for the
composite operator, to a good approximation it is just a Gaussian
theory, with a controlled expansion in $1\over N$ giving the
interactions.

\subsection{Introducing an elementary scalar}

The theory we would like to study is rewritten in terms of an
elementary scalar $\sigma$ and a Lagrange multiplier field
$\lambda$. We are assuming that a parameter $1/N$ can be introduced as
above, but for simplicity we do not write it explicitly.
\be
Z[j]=\int {\cal D}\phi ~e^{i \int _x~[\hf \phi \Box \phi + j\phi^2) ]}=\int {\cal D}\phi\int {\cal D}\sigma \int {\cal D}\lambda~  e^{i \int _x~[\hf \phi \Box \phi + \lambda (\sigma - \phi^2) ]+j\sigma}
\ee
We introduce a cutoff and rotate to Euclidean space. We write $\Box =
\Delta^{-1}$ and then write $\Delta= \Delta_l +\Delta_h$ and
$\phi=\phi_h+\phi_l$ to get:
\be
Z[j]=\int {\cal D}\phi _l \int {\cal D}\phi _h \int {\cal D}\sigma
\int {\cal D}\lambda~  e^{- \int _x~[\hf \phi_l\Delta_l^{-1}
  \phi_l+\hf \phi_h \Delta_h^{-1}\phi_h - \lambda (\sigma -
  (\phi_l+\phi_h))^2) ]+j\sigma} 
\ee
Do the $\phi_h$ integral:
\begin{equation}
\int {\cal D} \phi_h e^{-\int_x [\hf \phi_h( \Delta_h^{-1} +
   2\lambda)\phi_h +2\lambda \phi_l \phi_h]}\nn\\
= e^{-\hf Tr \ln (\Delta_h^{-1}+2\lambda) + 2\lambda \phi_l
  [(\Delta_h^{-1}+2\lambda)]^{-1}\lambda \phi_l}
\end{equation}

For simplicity we set $\phi_l=0$ and expand the logarithm to get
\be
e^{\hf Tr \ln \Delta_h - [Tr \Delta_h \lambda -\lambda \sigma] +
  Tr[\lambda \Delta_h^2 \lambda] + \int_x~j\sigma} 
\ee
Let us call $Tr\Delta_h =\sigma_0$ and write $\sigma=
\sigma_0+\sigma'$ to get
\be
e^{\hf Tr \ln \Delta_h  + Tr[\lambda \Delta_h^2 \lambda]
  +\int_x~[\lambda \sigma'+ j(\sigma'+\sigma_0)] +O(1/N)}\,. 
\ee
We now perform the $\lambda$ integral to get:
\be
e^{-\int_x~\hf \sigma' {1\over 2(\Delta_h^2)}\sigma' + j\sigma' +j\sigma_0}\,,
\ee
where we have dropped field independent parts.  This is now of the
form of an elementary scalar field coupled to a source and one can
apply the ERG analysis to this system. The high energy propagator for
the scalar field is $\tilde \Delta_h \equiv 2 \Delta_h^2$ as expected
from the two-point function for $\phi^2$. The ERG equation for $W[j]$
is as given in \eqref{ERGW} with $\tilde \Delta_h$ replacing
$\Delta_h$. The rest of the analysis then goes through as in Section
2.

\section{Nontrivial Fixed  Point}
\label{secFP}

The Polchinski ERG equation can be modified in order to discuss
perturbations about a nontrivial fixed point --- as is typically the
case in the AdS/CFT context, where the fixed point is assumed to be
some nontrivial and also strongly coupled CFT --- such as
suspersymmetric N=4 Yang-Mills. Perturbations are added to this action
and one is interested in two- and higher-point correlation functions.

Towards this end we start with some general background action and
perturb around it. We will be interested in the case where the
background is a fixed point action. But the fixed point nature is not
essential --- it just simplifies the time dependence.

%\subsection{Perturbing about a background action}

Let us start with a background action which we would like to perturb
by adding $S_1$.  Consider an action in $d=0$ for simplicity. Let the
background action, without any loss of generality, be
\[
S_{FP}= \hf x^2 G^{-1} + S_0(x)\,.
\]
Let the perturbation be $S_1$ so that the full action is
\[
S= \hf x^2 G^{-1} + S_0(x) + S_1 (x)\,.
\]
Then in our earlier notation, the ``wave functions'' are given by
\[
\psi = e^{-S}=e^{-[\hf x^2 G^{-1} + S_0(x,t) + S_1 (x,t)]}\,,\qquad
\psi'=e^{-[ S_0(x,t) + S_1 (x,t)]}\,. 
\]
Polchinski's equation is
\be	
 {\p \psi'\over \p t} = -\hf \dot G {\p^2 \psi'\over \p x^2}
\ee
What is special is that $S_0$ by itself satisfies Polchinski's
equation --- eventually it will be taken to be a fixed point solution.
Thus we have the following two equations: 
\be \label{s0} 
{\p S_0\over
  \p t} =\hf \dot G [-{\p^2S_0\over \p x^2} + ({\p S_0\over \p x})^2 ]
\ee
and 
\be \label{s1} 
{\p S_0\over \p t} +{\p S_1\over \p t}=\hf \dot
G [-{\p^2S_0\over \p x^2} + ({\p S_0\over \p x})^2 - {\p^2S_1\over \p
  x^2} + ({\p S_1\over \p x})^2 + 2({\p S_0\over \p x}) ({\p S_1\over
  \p x})] \,.
\ee
Subtracting \eqref{s0} from \eqref{s1} we get
\be \label{s2}
{\p S_1\over \p t}=\hf \dot G [- {\p^2S_1\over \p x^2} + ({\p S_1\over
  \p x})^2 + 2({\p S_0\over \p x}) ({\p S_1\over \p x})]\,. 
\ee
Since $S_0$ is a solution of \eqref{s0}, its form is ({\em in
  principle}) known as a function of time. In the case that $S_0$ is
chosen to be a fixed point solution, its time dependence can be
specified very easily: expressed in terms of {\em rescaled and
  dimensionless} variables it has no time dependence. This is
equivalent to saying that the dimensionless couplings are constant in
RG-time, $t$, i.e., they have vanishing beta functions. One can work
backwards and determine the exact $t$-dependence in terms of the
original variables. This is worked out in Appendix \ref{sec:D}.

\eqref{s2} can be used to define a modified Hamiltonian evolution
equation for the wave function $\psi''=e^{-S_1(x,t)}$:
\be \label{s3}
{\p \over \p t}\psi '' = -\hf \dot G [{\p ^2\over \p x^2} - 2 ({\p S_0\over \p x}){\p \over \p x}]\psi ''\,.
\ee
This can be generalized to $d$ dimensions easily, by replacing $x$ by
$x(p), x(-p)$ in appropriate places and integrating over $p$.  The
functional integral representation of this evolution can be obtained
using the $d+1$ dimensional action
\begin{eqnarray}
S &=& -\hf \int dt~\int {d^dp\over (2\pi)^d}~ \left[{1\over \dot G(p)}
  ({dx(p,t)\over dt})({dx(-p,t)\over dt}) \right.\nn\\
&&\left. \quad + \dot G(p) ({\delta S_0[x(p,t),t]\over \delta
    x(p,t)})({\delta S_0[x(p,t),t]\over \delta x(-p,t)})
+ 2 \dot x(p,t)({\delta S_0\over \delta x(p,t)})\right]\,.
\label{action}
\end{eqnarray}
This is derived in Appendix \ref{secA}.  In the above equation
$S_0[x(p,t),t]\equiv S_0[x(p),t]$ is a solution of \eqref{s0}. Thus in
general the bulk equation of motion is expected to be nonlinear,
which is indeed the case in AdS/CFT examples.

The last term in the action is a total derivative and the solution for
$\psi ''(x_f,T)$ can be written as
\be 
\psi ''(x_f(p),T)=e^{S_0[x_f(p),T]}\int {\cal
  D}x_i(p)\int_{x(p,0)=x_i(p)} {\cal
  D}x(p,t)e^{-S[x(p,t)]}e^{-S_0[x_i(p),0]-S_1[x_i(p),0]} \,,
\ee
where 
\[
S[x(p,t)]=\int dt ~\int {d^dp\over (2\pi)^d}~ [{1\over \dot G(p)}
({dx(p,t)\over dt})({dx(-p,t)\over dt}) + \dot G(p) ({\delta
  S_0[x(p,t),t]\over \delta x(p,t)})({\delta S_0[x(p,t),t]\over \delta
  x(-p,t)})]\,. 
\]

Equivalently the full solution can be written as
\begin{eqnarray}
&&\psi'(x_f(p),T) = e^{-S_0[x_f(p),T]}\psi
''(x_f(p),T)=e^{-S_0[x_f(p),T]-S_1[x_f(p),T]} \nn\\
&&= \int {\cal D} x_i(p)\int_{x(p,0)=x_i(p)} {\cal
  D}x(p,t)e^{-S[x(p,t)]}e^{-S_0[x_i(p),0]-S_1[x_i(p),0]} \,.
\end{eqnarray}
It is important to note that in the above expression $S_0[x(p,t),t]$
is a {\em known} function of time (unlike $S_1$), that has already
been obtained by solving the ERG equation \eqref{s0}. It is usually
taken to be the fixed point action that defines the boundary CFT. But
one can also imagine more general choices.

\section{Summary and Conclusions}

Let us summarize what was done in this paper:
\begin{enumerate}
\item We started with an ERG equation for the Wilson action and rewrote the
evolution operator as a functional integral, with ``time'' being the
ERG time $t=\ln {\Lambda_0\over \Lambda}$. Thus a d-dimensional theory
gives us a $d+1$ dimensional action. In this paper the $d$ dimensional
theory we consider describes a free scalar field $x(p)$ with a gaussian Wilson action
\[
S_\Lambda = \hf \int d^dp~ x (p)G(p)^{-1} x(-p)
\]
\item We then made a change of variables in the $d+1$ dimensional
  action (local on a scale $\Lambda$) of the form
  $x(p,t)=y(p,t)f(p,t)$, where $x$ is the original field variable and
  $y$ is the new variable. The function $f$ is chosen so that the
  action for $y$ is that of a scalar field in $AdS_{d+1}$. This
  requirement puts a constraint on $G$ and it turns out to be a ratio
  of Bessel functions, $K_m,I_m$ as given in \eqref{G}. This change of
  variables can be thought of as a coarse graining, with $y$ being the
  low energy field variable.
\item If we evaluate the on-shell action for $y$ we recover the result
  of evolving by the ERG evolution operator - low energy $d$
  dimensional action and Green function $G$ {\em in the limit
    $p\to 0$}. To recover the action fully, we need to add a boundary
  term involving $f$.  This is thus reminiscent of the AdS/CFT
  prescription for evaluating two-point functions. The difference is
  that we derive it mathematically starting from an ERG equation. It
  does not depend on the AdS/CFT conjecture.
\item We show that the boundary theory for the above special choice of
  $G$ (with a UV cutoff) has scale and conformal invariance induced by
  the $SO(d,2)$ isometry of the bulk theory. This involves a
  transformation of the cutoff $\Lambda$ (which is now the radial
  coordinate of $AdS$). More generally we know from \cite{Rosten1,
    Sonoda} that for arbitrary $G$ and for any fixed point one can
  define suitable scale and conformal transformations that leave the
  Wilson action invariant.
\item To apply the above formalism in the holographic context, one has
  to consider perturbing the theory with composite operators by adding
  terms of the form $J\phi^2$. We invoked a large N expansion to
  replace the composite operator by an elementary scalar (Section
  5). An ERG equation can then be written for $W_\Lambda[J]$ and the
  above procedure can be repeated. The leading term in the $1\over N$
  expansion gives the two-point function for the composite
  operator. Thus we are able to recover the AdS/CFT prescription for
  the two-point function, {\em in the presence of a UV cutoff}.
\item The entire discussion in this paper involved mathematical
  manipulations of Wilson's ERG and did not invoke the AdS/CFT
  conjecture. The final calculations can be interpreted, if one so
  desires, as ``holographic RG''.
\end{enumerate}

There are many open questions. We have two scales $\Lambda_0$ (cutoff
of the bare theory) and $\Lambda$. The usual AdS/CFT calculations take
$\Lambda_0\to \infty$ and deal with continuum theories. Our aim is to
keep $\Lambda_0$ finite. Some of our calculations involve keeping
$\Lambda_0$ finite and $\Lambda=0$. In Section 4 we have a discussion
of the case of finite $\Lambda$ but $\Lambda_0=\infty$ (showing that
the resulting Green function is analytic at $p=0$).  The full mapping
in the case when both are finite needs to be worked out. Also most of
our calculations involved the Gaussian fixed point. Only a preliminary
discussion of the nontrivial fixed point has been given (Section
6). We have not discussed interactions.  In the ERG the interactions
are included in the initial $S_B$. In holographic RG it comes out of
nonlinear terms in the evolution.  The modification of the evolution
kernel of the ERG is not done in usual ERG theory. However it is
reminiscent of what is done in ``MERA''\cite{Vidal}. This is worth
understanding.

We hope to return to these questions soon. 

\appendix

\section{Action Functional for Polchinski ERG using Canonical Formalism}
\label{secA}

Finally the same action can be obtained by converting everything in \eqref{Wilson} to Minkowski time (including $\dot G$)
Thus
\[ {dx\over dt_E}= -i{dx\over dt_M},~~~~~{dG\over dt_E}=-i{dG\over dt_M}
\]
to obtain
\be  
-i{\partial \psi'\over \partial t_M} = +{i\over 2} \dot G{\p \over \p
  y}({\p\over \p y} + 2G^{-1}y)\psi' -{i\over 2}\dot G G^{-1}\psi' 
\ee
So the Hamiltonian is (neglecting field independent terms):
\[
H= {i\over 2} \dot G[p^2-ip 2yG^{-1}]
\]
Using
\[
\dot y = {\p H\over \p p}=i\dot G p + {\dot G\over G} y ~\implies p= -i ({\dot y \over \dot G}- {y\over G})
\]
and 
\[
L_M= p\dot y -H\,,
\]
this gives
\[
L_M= i {\dot G\over 2}  ({\dot y\over \dot G}-{y\over G})^2 \implies L_E= -{\dot G\over 2}  ({\dot y\over \dot G}-{y\over G})^2\,.
\]
Then the Euclidean path integral becomes
\[
e^{\hf \int dt ~{\dot G\over 2}({\dot y\over \dot G}-{y\over G})^2}=e^{\hf \int dt ~{1\over 2\dot G}({\dot y}-{y\dot G \over G})^2}
\]
as obtained before.

\subsection*{Action for nontrivial fixed point}

As in the example above we rotate to Minkowski space: $it_M=t_E$.
Hence,
\[
e^{iS_M}=e^{i\int dt_M (-V)} =e^{-\int dt_E V}=e^{-S_E}
\]
Thus, $S_E=-iS_M$.  Thus in the above case  $S_{0}=-iS_{0M}$.
Write \eqref{s3} as
\[
{\p\over \p t_E} =-\hf \dot G [{\p ^2\over \p x^2} - 2 ({\p S_0\over
  \p x}){\p \over \p x} \implies 
{\p\over \p G}=-{\p\over \p \tau_E}=i{\p\over \p \tau_M}
=-\hf [{\p^2\over \p x^2} - 2 ({\p S_0\over \p x}){\p \over \p x}]=H 
\]
Writing in terms of $P= -i{\p\over \p x}$ we get finally for the Hamiltonian in Minkowski space-time,
\be
H= \hf[P^2 + 2 {\p S_{0M}\over \p x}P]
\ee
This gives 
\[
\dot x = P + {\p S_{0M}\over \p x}
\]
and one can obtain using $L=x\dot P -H$ an action
\be 
S_M=i\hf \int d\tau_M [\dot x^2 - 2\dot x {\p S_{0M}\over \p x} + {\p S_{0M}\over \p x}^2]
\ee
In Euclidean space this becomes
\[
\hf \int d\tau_E[\dot x^2 - 2\dot x {\p S_{0M}\over \p x} + {\p S_{0M}\over \p x}^2]
\]
Rewriting in terms of $G=-\tau_E$ we get
\be 
S_E=-\hf \int dt [{1\over \dot G}\dot x^2 - 2\dot x {\p S_{0M}\over \p x} +\dot G{\p S_{0M}\over \p x}^2]
\ee
Introducing $x(p)$ and integrating over $p$ we get \eqref{action}.

\section{Conformal Invariance}
\label{sec:B}
\subsection{Generators}

$AdS_3$ space is parametrized by: (Generalization to higher spaces is trivial)
\be \label{ads}
-(y^0)^2+(y^1)^2+(y^2)^2-(y^4)^2=-1
\ee
Metric is thus {-++-}. $y^4=-y_4, y^0=-y_0$

To ensure covariant notation we define the generators: 
\be
J_{\mu\nu}= y_\mu {\p\over \p y^\nu}- y_\nu {\p\over \p y^\mu}
\ee
\begin{subequations}
\br
T_{01}&=&y_0{\p\over \p y^1}-y_1{\p\over \p y^0} \implies 
\left\lbrace\begin{array}{c@{~=~}l}
\delta y^1 & \epsilon y_0 = -\epsilon y^0 \\
\delta y^0 & -\epsilon y_1 = -\epsilon y^1 
\end{array}\right.\\
T_{02}&=&y_0{\p\over \p y^2}-y_2{\p\over \p y^0} \implies 
\left\lbrace\begin{array}{c@{~=~}l}
\delta y^2 & \epsilon y_0 = -\epsilon y^0 \\
\delta y^0 & -\epsilon y_2 = -\epsilon y^2
\end{array}\right.\\
T_{04}&=&y_0{\p\over \p y^4}-y_4{\p\over \p y^0} \implies 
\left\lbrace
\begin{array}{c@{~=~}l}
\delta y^4  & \epsilon y_0 = -\epsilon y^0 \\ 
\delta y^0 & -\epsilon y_4 = +\epsilon y^4 
\end{array}\right.\\
T_{12}&=& y_1{\p\over \p y^2}-y_2{\p\over \p y^1} \implies 
\left\lbrace\begin{array}{c@{~=~}l}
\delta y^2 & \epsilon y_1 = \epsilon y^1 \\ 
\delta y^1 & -\epsilon y_2 = -\epsilon y^2
\end{array}\right.\\
T_{14}&=&y_1{\p\over \p y^4}-y_4{\p\over \p y^1} \implies 
\left\lbrace\begin{array}{c@{~=~}l}
\delta y^4 & \epsilon y_1 = \epsilon y^1 \\ 
\delta y^1 & -\epsilon y_4 = +\epsilon y^4 
\end{array}\right.
\er
\end{subequations}
\be
U=y^2+y^4;  V=y^2-y^4
\ee
\be	\label{5}
UV =-1 + (y^0)^2 -(y^1)^2 \implies V= {-1-y_\mu y^\mu\over U}
\ee
One can check that
\begin{subequations}
\begin{eqnarray}
T_{12}-T_{14} &:& \delta y^1= -\epsilon U\\
T_{12}+T_{14} &:& \delta y^1= -\epsilon V\\
T_{02}-T_{04} &:& \delta y^0= -\epsilon U\\
T_{02}+T_{04}&:& \delta y^2= -\epsilon V
\end{eqnarray}
\end{subequations}
We will see that $T_{12}-T_{14}=P_1$ (translation) and
$T_{12}+T_{14}=C_1$ (special conformal transformation).  Similarly
$T_{02}-T_{04}=P_0$ and $T_{02}+T_{04}=C_0$. Also
\begin{eqnarray}
P_1 &:& \delta U=0, \,\delta V= 2\epsilon y^1\\
C_1 &:& \delta U=2\epsilon y^1; \,\delta V=0
\end{eqnarray}
Write this covariantly:
\be
C_\mu: \delta U= 2\epsilon y_\mu; ~~~~ \delta y^\nu= -\epsilon V \delta _\mu^\nu
\ee
Poincare patch coordinates:
\be 
x^\mu = {y^\mu\over U}~;~~~~ z= {1\over U}
\ee
\br
C_\nu : \delta  x^\mu &=& {\delta y^\mu\over U} -{y^\mu \delta U\over U^2}\\ \nonumber
&=&-{\epsilon V \delta _\nu ^\mu\over U} - {y^\mu 2 \epsilon y_\nu\over U^2} \\ \nonumber
&=&+\epsilon(z^2+x_\rho x^\rho) \delta_\nu^\mu - 2\epsilon x_\nu x^\mu
\er
Thus
\be
C_\nu = (z^2+x_\rho x^\rho){\p\over \p x^\nu} - 2 x_\nu x^\rho {\p\over \p x^\rho}
\ee
Acting on $z$:
\[
C_\nu z = C_\nu {1\over U} : \delta {1\over U} = {-2\epsilon y_\nu \over U^2} = -2\epsilon z x_\mu
\]
Thus
\be
C_\nu = -2x_\mu z{d\over dz}
\ee
Thus combining everything:
\be
C_\nu = (z^2+x_\rho x^\rho){\p\over \p x^\nu} - 2 x_\nu( x^\rho {\p\over \p x^\rho}+z{d\over dz})
\ee
Furthermore including a scaling dimension $\Delta$ gives the final form:
\be    \label{specconf}
C_\nu = (z^2+x_\rho x^\rho){\p\over \p x^\nu} - 2 x_\nu( x^\rho {\p\over \p x^\rho}+z{d\over dz}+\Delta)
\ee
It is also clear that $P_\mu$ are translations:
\[
P_\mu : \delta x^\nu = -\epsilon \delta_\mu^\nu
\]

\subsection*{Dilatations:}

\begin{eqnarray*}
&&T_{24}= y_2{\p\over \p y^4} -y_4{\p\over \p y^2} ~~:\delta y^2 =
   -\epsilon y_4 = + \epsilon y^4;~~\delta y^4 = \epsilon y_2 =
   \epsilon y^2\\ 
&&\implies \delta x^1 = -\epsilon x^1;~~~\delta x^0 = -\epsilon
   x^0;~~~\delta z = -{\delta U\over U^2}=-\epsilon z\\ 
&&[K_1,P_1]=[T_{12}+T_{14},T_{12}-T_{14}]= T_{24} = D = -x^\mu
   {\p\over \p x^\mu} - z{\p\over \p z} 
\end{eqnarray*}
Again, including a scaling dimension $\Delta$ this becomes:
\be   \label{Dil}
D = -x^\mu {\p\over \p x^\mu} - z{\p\over \p z}-\Delta
\ee

\subsection*{Momentum Space}

We give the generators in momentum space.
We write $C_\mu = C^1_\mu + C^2_\mu +C^3_\mu$ where
\begin{subequations}
\label{cftr}
\begin{eqnarray}
C_\mu^1 &=& 2(d-\Delta) {\p\over \p p^\mu } + 2 p^\rho {\p^2\over \p
            p^\rho \p p^\mu} -p_\mu{\p^2\over \p p^\rho \p p_\rho}\\ 
C_\mu^2 &=& -2z {\p^2\over \p z \p p^\mu} \\
C^3_\mu &=& (z^2-z'^2)p_\mu
\end{eqnarray}
\end{subequations}
$C^1_\mu$ has been calculated to be (acting on functions of $p^2$)
\[
C_\mu^1 = p_\mu[{d^2\over dp^2}+{d+1-2\Delta\over p} {d\over dp}]
\]
This has been used in \eqref{ct}.

Furthermore it is illuminating to combine the terms involving $z { d\over dz}$ in $C_\mu^1+C_\mu^2+C_\mu^3$. We get the combination
\be
2(d-\Delta) {\p\over \p p^\mu } -2z {\p^2\over \p z \p p^\mu} = -2(\Delta-d + z { d\over dz}){\p\over \p p^\mu }
\ee
In this expression $\Delta-d$ is the dimension of the field $\phi(p)$ and $\Delta $ is the dimension of $\phi(x)$. This is the same combination that occurs in dilatations:
\be
\delta_{dil} \phi =- [\Delta -d+ z { d\over dz}- p^\nu{\p\over \p p^\nu}]\phi(p)
\ee

\subsection{Invariance of Action}
We consider the action 
\be
S_1=\hf \int d^dx~\int_{\epsilon_1}^{\epsilon_2} {dz\over z}~z^{-d+2} [\p_z \phi (x,z)\p_z \phi(x,z) + \p_i \phi(x,z)\p_i \phi(x,z)]
\ee
and the transformation given by \eqref{Dil} with $\Delta=0$:
\[
\delta \phi = D\phi= -(x^i{\p\over \p x^i}+ z{\p\over \p z})\phi(z,x)
\]
It is easy to see that we are left with a total derivative
\[
-\hf \int d^dx~\int_{\epsilon_1}^{\epsilon_2} dz~\p_z [ z^{-d+2} (\p_z\phi \p_z \phi + \p_k \phi \p_k \phi )] 
\]
which gives the boundary term \eqref{d-b} of Section \ref{2}

Similarly under the variation $C_i$ \eqref{specconf} (with $\Delta=0$)
\[
\delta \phi=C_i \phi = [(z^2+x_k x^k){\p\over \p x^i} - 2 x_i( x^k {\p\over \p x^k}+z{d\over dz})]\phi(z,x)
\]
we get the total derivative:
\[
- \int d^dx~\int_{\epsilon_1}^{\epsilon_2} dz~\p_z [ z^{-d+2}x_i (\p_z\phi \p_z \phi + \p_k \phi \p_k \phi )]
\]
we get the boundary term \eqref{sc-b} of Section \ref{2}. These are canceled by the variations of $\epsilon$ given in Section \ref{2}.

\section{ERG for $W_\Lambda[J]$}
\label{sec:C}

Given a bare action
\begin{equation}
S_B [\phi] = \frac{1}{2} \phi \Delta^{-1} \phi + S_{B,I} [\phi]\,,
\end{equation}
the generating functional of correlation functions is defined by
\begin{equation}
Z[J] \equiv e^{W_B [J]} = \int \mathcal{D} \phi\, e^{- S_B [\phi] + J
  \phi}\,.
\end{equation}
We split the bare propagator into high and low energy propagators:
\begin{equation}
\Delta = \Delta_h + \Delta_l = \frac{e^{- \frac{p^2}{\Lambda_0^2}}}{p^2}\,,
\end{equation}
where we choose
\begin{equation}
\left\lbrace
\begin{array}{c@{~=~}l}
\Delta_l & \frac{e^{- \frac{p^2}{\Lambda^2}}}{p^2}\,,\\
\Delta_h & \frac{e^{- \frac{p^2}{\Lambda_0^2}}- e^{-
           \frac{p^2}{\Lambda^2}}}{p^2}\,.
\end{array}\right.\quad (\Lambda_0 > \Lambda)
\end{equation}
The interaction part of the Wilson action at cutoff $\Lambda$
is defined by
\begin{equation}
e^{- S_{\Lambda, I} [\phi_l]}
\equiv \int \mathcal{D} \phi_h\, e^{- \frac{1}{2} \phi_h
  \Delta_h^{-1} \phi_h - S_{B,I} [\phi_h+\phi_l]}\,,
\end{equation}
and the total Wilson action by
\begin{equation}
S_\Lambda [\phi] \equiv \frac{1}{2} \phi \Delta_l^{-1} \phi +
S_{\Lambda, I} [\phi]\,.
\end{equation}

We now follow \cite{MorrisERG} and define
\begin{equation}
W_\Lambda [J] \equiv \frac{1}{2} J \Delta_h J - S_{\Lambda, I}
\left[ \Delta_h J \right]\,.
\end{equation}
Let us show
\begin{equation}
e^{W_B [J]} = \int \mathcal{D} J'\, \exp \left( W_\Lambda [J'] -
  \frac{1}{2} (J'-J) \frac{\Delta \Delta_h}{\Delta_l} (J'-J)\right)\,,
\label{WB-WLambda}
\end{equation}
where
\begin{equation}
\frac{\Delta_l}{\Delta \Delta_h} = \frac{\Delta - \Delta_h}{\Delta
  \Delta_h} = \frac{1}{\Delta_h} - \frac{1}{\Delta}\,.\label{AppendixC-Delta}
\end{equation}
We verify (\ref{WB-WLambda}) by computing its right-hand side:
\begin{eqnarray}
(\mathrm{RHS}) &=& \int \mathcal{D} J'\, \exp \left( \frac{1}{2} J'
                   \Delta_h J' - S_{\Lambda, I} \left[ \Delta_h
                   J' \right] \right.\nn\\
&&\left.\qquad\qquad- \frac{1}{2} (J'-J) \frac{\Delta
                   \Delta_h}{\Delta_l} (J'-J) \right)\nn\\
&=& \int \mathcal{D} J'\, \mathcal{D} \phi'\,  \exp \left( \frac{1}{2} J'
                   \Delta_h J' - \frac{1}{2} \phi' \Delta_h^{-1}
    \phi'\right.\nn\\
&&\left.\qquad - S_{B,I} \left[ \phi' + \Delta_h J' \right]
  -   \frac{1}{2} (J'-J) \Delta \Delta_h \Delta_l^{-1}
                   (J'-J) \right)\,.
\end{eqnarray}
Shifting $\phi'$ by $\Delta_h J'$, we obtain
\begin{eqnarray}
(\mathrm{RHS}) &=&\int  \mathcal{D} \phi' \mathcal{D} J'\,\,  \exp \left(
    \frac{1}{2} J' \Delta_h J' - \frac{1}{2} \left( \phi' -
    \Delta_h J' \right) \Delta_h^{-1} \left(\phi' - \Delta_h J'
    \right) \right.\nn\\
&&\qquad\left.  - S_{B,I} \left[ \phi' \right]
+   \frac{1}{2} (J'-J)
   \Delta \Delta_h \Delta_l^{-1}
                   (J'-J) \right)\nn\\
&=&  \int  \mathcal{D} \phi' \, e^{-S_B [\phi'] + J \phi'} \mathcal{D} J'\,
\exp \left(  (J'-J) \phi' \right.\nn\\
&&\qquad\left. - \frac{1}{2}  (J'-J)
   \Delta \Delta_h \Delta_l^{-1} (J'-J)  
+ \frac{1}{2} \phi' (\Delta^{-1}-\Delta_h^{-1}) \phi' \right) \,.
\end{eqnarray}
Using (\ref{AppendixC-Delta}) and shifting $J'$ by
$J-\frac{\Delta_l}{\Delta\Delta_h} \phi'$, we obtain the desired result
\begin{eqnarray}
(\mathrm{RHS}) &=&  \int  \mathcal{D} \phi' \, e^{S_B [\phi'] + J
  \phi'} \int \mathcal{D} J'\, 
\exp \left( - \frac{1}{2} J' \frac{\Delta \Delta_h }{\Delta_l}
    J' \right)\nn\\
&=& e^{W_B [J]}\,.
\end{eqnarray}

We can rewrite (\ref{WB-WLambda}) as
\begin{equation}
e^{W_B [J]} = \exp \left( \frac{1}{2} \frac{\Delta_l}{\Delta\Delta_h}
  \frac{\partial^2}{(\partial J)^2} \right) e^{W_\Lambda [J]}\,.
\end{equation}
Inverting this, we obtain
\begin{equation}
e^{W_\Lambda [J]} = \exp \left( - \frac{1}{2} \frac{\Delta_l}{\Delta\Delta_h}
  \frac{\partial^2}{(\partial J)^2} \right) e^{W_B [J]}\,.
\end{equation}
This gives
\begin{equation}
- \Lambda \frac{\partial}{\partial \Lambda} e^{W_\Lambda [J]}
= \frac{1}{2}  \Lambda \frac{\partial}{\partial \Lambda}
\left(\frac{\Delta_l}{\Delta\Delta_h}\right) \frac{\partial^2}{(\partial J)^2}
e^{W_\Lambda [J]}\,.
\end{equation}
Using
\begin{equation}
\Lambda \frac{\partial}{\partial \Lambda}
\left(\frac{\Delta_l}{\Delta\Delta_h}\right) 
= \Lambda \frac{\partial}{\partial \Lambda} \left(\frac{1}{\Delta_h} -
  \frac{1}{\Delta}\right)
=  \Lambda \frac{\partial}{\partial \Lambda} \Delta_h^{-1}
\end{equation}
we finally obtain
\begin{equation}
- \Lambda \frac{\partial}{\partial \Lambda} W_\Lambda [J]
=  \frac{1}{2} \Lambda
  \frac{\partial \Delta_h^{-1}}{\partial \Lambda}
\left(  \frac{\partial W_\Lambda}{\partial J}
  \frac{\partial W_\Lambda}{\partial J} + \frac{\partial^2
    W_\Lambda}{\partial J^2} \right)\,.
\end{equation}

Incidentally, (\ref{WB-WLambda}) shows
\begin{equation}
W_B [J] = \lim_{\Lambda \to 0+} W_\Lambda [J]
\end{equation}
because $\Delta_l/\Delta_h \to 0$ as $\Lambda \to 0$.

\section{Rescaled Variables and $t$-dependence of Fixed Point Action}
\label{sec:D}

For a real scalar theory, the ERG differential equation in the
dimensionless convention (with rescaled momenta) is given by
\begin{eqnarray}
\partial_t S_t [\bar{\phi}] &=& \int_p \left(- p \cdot \partial_p \ln K(p) +
  \frac{d+2}{2} - \gamma + p \cdot \partial_p \right) \bar{\phi} (p) \cdot
\frac{\delta S_t}{\delta \bar{\phi} (p)}\nn\\
&& + \int_p \left( p \cdot \partial_p - 2
  \gamma \right) \frac{1-K(p)}{K(p)} \cdot \frac{K(p)^2}{p^2}\nn\\
&&\qquad\qquad \times\frac{1}{2}
\left\lbrace  \frac{\delta^2 S_t}{\delta \bar{\phi} (p) \delta
    \bar{\phi} (-p)} - \frac{\delta S_t}{\delta \bar{\phi} (-p)} \frac{\delta
    S_t}{\delta \bar{\phi} (p)} \right\rbrace\,,
\label{ERG-dimless}
\end{eqnarray}
where we use a shorthand notation
\begin{equation}
\int_p = \int \frac{d^d p}{(2\pi)^d}\,,
\end{equation}
and $K(p)$ is a cutoff function such as
\begin{equation}
K(p) = e^{- p^2}\,.
\end{equation}
Substituting the Gaussian fixed point
\begin{equation}
  S_G [\bar{\phi}] = \frac{1}{2} \int_p \frac{p^2}{K(p)} \bar{\phi}
  (p) \bar{\phi} (-p) 
\end{equation}
into the right-hand side ($\gamma = 0$), we obtain zero.  A
non-trivial fixed point would correspond to a positive anomalous
dimension $\gamma$.

To obtain the ERG differential equation in the dimensionful convention
(with no rescaling of momenta), we choose a reference momentum scale
$\mu$, and introduce a running momentum cutoff by
\begin{equation}
\Lambda = \mu\, e^{-t}\,.
\end{equation}
The dimensionful field is related to the dimensionless field by
\begin{equation}
\phi (p) = \Lambda^{-\frac{d+2}{2}} \bar{\phi} (p/\Lambda)\,.
\end{equation}
The action is given by
\begin{equation}
S_\Lambda [\phi] = S_t [\bar{\phi}]\,.
\end{equation}
Using
\begin{equation}
  - \Lambda \frac{\partial}{\partial \Lambda} \left(
    \Lambda^{-\frac{d+2}{2}} \bar{\phi} (p/\Lambda) \right) 
  = \left(\frac{d+2}{2} + p \cdot \partial_p \right) \phi (p)\,,
\label{transcription}
\end{equation}
we can rewrite (\ref{ERG-dimless}) as
\begin{eqnarray}
&&- \Lambda \frac{\partial}{\partial \Lambda} S_\Lambda [\phi]
= \int_p \left(\Lambda \frac{\partial}{\partial \Lambda} \ln
  K(p/\Lambda)-\gamma\right) \phi (p) \frac{\delta S_\Lambda
  [\phi]}{\delta \phi (p)}\label{ERG-dimful}\\
&& \quad + \int_p \left(- \Lambda \frac{\partial}{\partial \Lambda} -
  2 \gamma \right) \frac{1-K(p/\Lambda)}{K(p/\Lambda)} 
\cdot \frac{K(p/\Lambda)^2}{p^2} 
\cdot \frac{1}{2}
\left\lbrace \frac{\delta^2 S_\Lambda}{\delta
    \phi (p) \delta \phi (-p)} - \frac{\delta S_\Lambda}{\delta \phi
    (p)} \frac{\delta S_\Lambda}{\delta \phi (-p)}  \right\rbrace\,.
\nn
\end{eqnarray}

Even if $S_t$ is a fixed point with no $t$-dependence, the
corresponding $S_\Lambda$ depends on $\Lambda$.  
For example, the action of the Gaussian fixed point gives
\begin{equation}
S_\Lambda [\phi] = \frac{1}{2} \int_p \frac{p^2}{K(p/\Lambda)} \phi
(p) \phi (-p)\,.
\end{equation}
This satisfies (\ref{ERG-dimful}) with $\gamma = 0$.  More generally,
suppose a fixed point action is given as
\begin{equation}
S [\bar{\phi}] = \sum_{n=2}^\infty \frac{1}{n!} \int_{p_1, \cdots,
  p_n} \bar{\phi} (p_1) \cdots \bar{\phi} (p_n)\, (2\pi)^d
\delta^{(d)} (p_1+\cdots + p_n)\, S_n (p_1, \cdots, p_n)\,.
\end{equation}
Using (\ref{transcription}), we obtain the corresponding
$\Lambda$-dependent action as
\begin{eqnarray}
S_\Lambda [\phi] &=& \sum_{n=2}^\infty \frac{1}{n!} \int_{p_1, \cdots,
  p_n} \Lambda^{n \frac{d+2}{2}} \phi (p_1 \Lambda) \cdots \phi (p_n
\Lambda)\, (2\pi)^d
\delta^{(d)} (p_1+\cdots + p_n)\nn\\
&&\qquad \times S_n (p_1, \cdots, p_n)\nn\\
&=&  \sum_{n=2}^\infty \frac{1}{n!} \int_{p_1, \cdots,
  p_n}  \phi (p_1) \cdots \phi (p_n)\, (2\pi)^d \delta (p_1 + \cdots
p_n)\nn\\
&&\qquad \times \Lambda^{- n \frac{d-2}{2} + d}
\, S_n (p_1/\Lambda,
\cdots, p_n/\Lambda)\,.
\end{eqnarray}
If $S$ had $t$-dependence, $S_n$ would obtain explicit dependence on
$\Lambda$.

%%%%%%%%%%%%%%%%%%%%%%%%%%%%%%%%%%%%%%%%%

\section*{Acknowledgments}

BS would like to thanks Nemani Suryanarayana for some discussions.


\begin{thebibliography}{99}


\bibitem{Maldacena} 
  J.~M.~Maldacena,
  ``The Large N limit of superconformal field theories and supergravity,''
  Int.  J. Theor. Phys.  {\bf 38}, 1113 (1999)
  [Adv. Theor. Math. Phys.  {\bf 2}, 231 (1998)]
  doi:10.1023/A:1026654312961
  \texttt{arXiv:hep-th/9711200}.
  
\bibitem{Polyakov}  
S.~S.~Gubser, I.~R.~Klebanov, and A.~M.~Polyakov,
``Gauge theory correlators from non-critical string
theory,'' Phys. Lett. \textbf{B428} (1998) 105-114,
\texttt{arXiv:hep-th/9802109}.

\bibitem{Witten1} 
E.~Witten, ``Anti-de Sitter space and holography,'' 
Adv. Theor. Math. Phys. \textbf{2} (1998) 253-291,
\texttt{arXiv:hep-th/9802150}.
  
%\cite{Witten:1998zw}
\bibitem{Witten2} 
  E.~Witten,
  ``Anti-de Sitter space, thermal phase transition, and confinement in
  gauge theories,'' 
  Adv. Theor. Math. Phys.  {\bf 2}, 505 (1998)
  \texttt{arXiv:hep-th/9803131}.

\bibitem{BSLV} B.~Sathiapalan, ``Loop Variables, the Renormalization
  Group and Gauge Invariant Equations of Motion in String Field
  Theory,'' Nucl.\ Phys.\ {\bf B326}, 376 (1989).
  doi:10.1016/0550-3213(89)90137-5.
  
\bibitem{Witten3} 
  E.~Witten,
  ``Branes and the dynamics of QCD,''
  Nucl.\ Phys.\ {\bf B507}, 658 (1997)
  doi:10.1016/S0550-3213(97)00648-2
  \texttt{arXiv:hep-th/9706109}.
  

%\cite{Wilson:1973jj}
\bibitem{Wilson} 
  K.~G.~Wilson and J.~B.~Kogut,
  ``The Renormalization group and the epsilon expansion,''
  Phys. Rept. {\bf 12}, 75 (1974).
  doi:10.1016/0370-1573(74)90023-4
  %%CITATION = doi:10.1016/0370-1573(74)90023-4;%%

\bibitem{Wegner}
F.~J.~Wegner and A.~Houghton, ``Renormalization group
equation for critical phenomena,'' Phys.\ Rev.\ \textbf{A8} (1973)
401-412.

\bibitem{Wilson2} 
K.~G.~Wilson, ``The renormalization group and critical
phenomena,''  Rev. Mod. Phys. \textbf{55} (1983) 583-600.

\bibitem{Polchinski} 
  J.~Polchinski,
  ``Renormalization and Effective Lagrangians,''
  Nucl. Phys. {\bf B231}, 269 (1984).
  doi:10.1016/0550-3213(84)90287-6
  
\bibitem{MorrisERG} 
  T.~R.~Morris,
  ``The Exact renormalization group and approximate solutions,''
  Int.\ J.\ Mod.\ Phys.\ A {\bf 9}, 2411 (1994)
  doi:10.1142/S0217751X94000972
  \texttt{arXiv:hep-ph/9308265}.
  
\bibitem{Becchi}
C.~Becchi, ``On the construction of renormalized gauge theories using
                        renormalization group techniques,''
\texttt{arXiv:hep-th/9607188}
%%CITATION = HEP-TH/9607188;%%

\bibitem{Bagnuls1} 
  C.~Bagnuls and C.~Bervillier,
  ``Exact renormalization group equations and the field theoretical approach to critical phenomena,''
  Int.\ J.\ Mod.\ Phys.\ A {\bf 16}, 1825 (2001)
  doi:10.1142/S0217751X01004505
  \texttt{hep-th/0101110}.
  %%CITATION = doi:10.1142/S0217751X01004505;%%
  %11 citations counted in INSPIRE as of 02 Jun 2017

%\cite{Bagnuls:2000ae}
\bibitem{Bagnuls2} 
  C.~Bagnuls and C.~Bervillier,
  ``Exact renormalization group equations. An Introductory review,''
  Phys.\ Rept.\  {\bf 348}, 91 (2001)
  doi:10.1016/S0370-1573(00)00137-X
  \texttt{hep-th/0002034}.
  %
\bibitem{Igarashi} 
  Y.~Igarashi, K.~Itoh, and H.~Sonoda,
  ``Realization of Symmetry in the ERG Approach to Quantum Field Theory,''
  Prog.\ Theor.\ Phys.\ Suppl.\  {\bf 181}, 1 (2010)
  doi:10.1143/PTPS.181.1
  \texttt{arXiv:0909.0327 [hep-th]}.

\bibitem{Rosten:2010vm}
O.~J.~Rosten, ``Fundamentals of the Exact Renormalization Group,''
Phys.\ Rep.\ \textbf{511} (2012)177-272,
\texttt{arXiv:1003.1366 [hep-th]}.
  
\bibitem{Akhmedov}
E.~T.~Akhmedov, ``A Remark on the AdS / CFT
correspondence and the renormalization group flow,''
Phys. Lett.  \textbf{B442} (1998) 152-158,
\texttt{arXiv:hep-th/9806217 [hep-th]}.

\bibitem{Alvarez} 
E.~Alvarez and C.~Gomez, ``Geometric holography, the
renormalization group and the c theorem,'' Nucl.Phys.
\textbf{B541} (1999) 441-460, \texttt{arXiv:hep-th/9807226
[hep-th]}.

\bibitem{Girardello} 
L.~Girardello, M.~Petrini, M.~Porrati, and A.~Zaffaroni,
``Novel local CFT and exact results on perturbations of
N=4 superYang Mills from AdS dynamics,'' JHEP
\textbf{9812} (1998) 022, \texttt{arXiv:hep-th/9810126 [hep-th]}.

\bibitem{Distler}
J.~Distler and F.~Zamora, 
``Nonsupersymmetric
conformal field theories from stable anti-de Sitter
spaces,'' Adv.  Theor.  Math.  Phys. \textbf{2} (1999) 1405-1439,
\texttt{arXiv:hep-th/9810206 [hep-th]}.

\bibitem{Kraus} 
V.~Balasubramanian and P.~Kraus, ``Space-time and the
holographic renormalization group,'' Phys. Rev. Lett. \textbf{83}
(1999) 3605-3608, \texttt{arXiv:hep-th/9903190 [hep-th]}.

\bibitem{Warner} 
D.~Freedman, S.~Gubser, K.~Pilch, and N.~Warner,
``Renormalization group flows from holography
supersymmetry and a c theorem,''
Adv. Theor. Math. Phys. \textbf{3} (1999) 363-417,
\texttt{arXiv:hep-th/9904017 [hep-th]}.

\bibitem{Verlinde} 
J.~de~Boer, E.~P.~Verlinde, and H.~L.~Verlinde, ``On the
holographic renormalization group,'' JHEP \textbf{08} (2000)
003, \texttt{arXiv:hep-th/9912012}.

\bibitem{Boer} J.~de~Boer, 
``The Holographic renormalization group,''
Fortsch.~Phys. \textbf{49} (2001) 339-358,
\texttt{arXiv:hep-th/0101026 [hep-th]}.

  %%CITATION = doi:10.1016/0550-3213(84)90287-6;%%  %\cite{Faulkner:2010jy}
\bibitem{Faulkner} 
  T.~Faulkner, H.~Liu, and M.~Rangamani,
  ``Integrating out geometry: Holographic Wilsonian RG and the membrane paradigm,''
  JHEP {\bf 1108}, 051 (2011)
  doi:10.1007/JHEP08(2011)051
\texttt{arXiv:1010.4036 [hep-th]}.

\bibitem{Lizana:2015hqb} 
  J.~M.~Lizana, T.~R.~Morris, and M.~Perez-Victoria,
  ``Holographic renormalisation group flows and renormalisation from a Wilsonian perspective,''
  JHEP {\bf 1603}, 198 (2016)
  doi:10.1007/JHEP03(2016)198
 \texttt{arXiv:1511.04432 [hep-th]}.
%\cite{Klebanov:1999tb}

\bibitem{Klebanov:1999tb} 
  I.~R.~Klebanov and E.~Witten,
  ``AdS / CFT correspondence and symmetry breaking,''
  Nucl. Phys. {\bf B556}, 89 (1999)
  doi:10.1016/S0550-3213(99)00387-9
  \texttt{arXiv:hep-th/9905104}.
 
 \bibitem{Heemskerk} 
  I.~Heemskerk and J.~Polchinski,
  ``Holographic and Wilsonian Renormalization Groups,''
  JHEP {\bf 1106}, 031 (2011)
  doi:10.1007/JHEP06(2011)031
\texttt{arXiv:1010.1264 [hep-th]}.   

\bibitem{Morris} 
  J.~M.~Lizana, T.~R.~Morris, and M.~Perez-Victoria,
  ``Holographic renormalisation group flows and renormalisation from a Wilsonian perspective,''
  JHEP {\bf 1603}, 198 (2016)
  doi:10.1007/JHEP03(2016)198
\texttt{arXiv:1511.04432 [hep-th]}.

%\cite{Bzowski:2015pba}
\bibitem{Bzowski:2015pba} 
  A.~Bzowski, P.~McFadden, and K.~Skenderis,
  ``Scalar 3-point functions in CFT: renormalisation, beta functions and anomalies,''
  JHEP {\bf 1603}, 066 (2016)
  doi:10.1007/JHEP03(2016)066
\texttt{arXiv:1510.08442 [hep-th]}.
  
  %\cite{deHaro:2000vlm}
\bibitem{deHaro:2000vlm} 
  S.~de Haro, S.~N.~Solodukhin, and K.~Skenderis,
  ``Holographic reconstruction of space-time and renormalization in the AdS / CFT correspondence,''
  Comm.\ Math.\ Phys.\  {\bf 217}, 595 (2001)
  doi:10.1007/s002200100381
\texttt{arXiv:hep-th/0002230}.
  
%\cite{Rosten:2014oja}
\bibitem{Rosten0}
O.~J.~Rosten,
``On Functional Representations of the Conformal Algebra,''
\texttt{arXiv:1411.2603 [hep-th]}
%%CITATION = ARXIV:1411.2603;%%

%\cite{Rosten:2016zap}
\bibitem{Rosten1} 
  O.~J.~Rosten,
  ``A Conformal Fixed-Point Equation for the Effective Average Action,''
\texttt{arXiv:1605.01729 [hep-th]}.
  %%CITATION = ARXIV:1605.01729;%%
  %1 citations counted in INSPIRE as of 24 May 2017

%\cite{Rosten:2016nmc}
\bibitem{Rosten2} 
  O.~J.~Rosten,
  ``A Wilsonian Energy-Momentum Tensor,''
  \texttt{arXiv:1605.01055 [hep-th]}.
  
  %\cite{Rosten:2017urs}
\bibitem{Rosten3} 
  O.~J.~Rosten,
  ``Wilsonian Ward Identities,''
  \texttt{arXiv:1705.05837 [hep-th]}.
  
  %\cite{Sonoda:2017zgl}
\bibitem{Sonoda} 
  H.~Sonoda,
  ``Conformal invariance for Wilson actions,''
  \texttt{arXiv:1705.01239 [hep-th]}.
  %%CITATION = ARXIV:1705.01239;%%

 \bibitem{Wetterich} 
  C.~Wetterich,
  ``Exact evolution equation for the effective potential,''
  Phys.\ Lett.\ {\bf B301}, 90 (1993).
  doi:10.1016/0370-2693(93)90726-X 
  
\bibitem{Polchinski2}  
J.~Polchinski,
``Scale and Conformal Invariance in Quantum Field Theory,''
Nucl. Phys. \textbf{B303} (1988) 226-236.
DOI: 10.1016/0550-3213(88)90179-4
  %\cite{Muck:1998rr}
\bibitem{Muck:1998rr} 
  W.~M\"uck and K.~S.~Viswanathan,
  ``Conformal field theory correlators from classical scalar field theory on AdS(d+1),''
  Phys.\ Rev.\ {\bf D58}, 041901 (1998)
  doi:10.1103/PhysRevD.58.041901
  \texttt{arXiv:hep-th/9804035}.
 
	
\bibitem{Oak} 
Prafulla Oak and B. ~Sathiapalan,  ``Exact Renormalization Group and Sine
Gordon Theory,'' \texttt{arXiv:1703.01591 [hep-th]}.

\bibitem{Vidal} 
  G.~Vidal,
  ``Entanglement Renormalization,''
  Phys.\ Rev.\ Lett.\  {\bf 99}, no. 22, 220405 (2007)
  doi:10.1103/PhysRevLett.99.220405
  [cond-mat/0512165].
\end{thebibliography}
\end{document}